\documentclass[12pt]{article}
\pdfoutput=1
\usepackage{jheppub}
\usepackage[utf8]{inputenc}
\usepackage{amsfonts,amsmath}

\usepackage{ifthen}
\usepackage{enumerate}
\usepackage{amsmath}
\usepackage{amssymb}
\usepackage[mathscr]{eucal}
\usepackage[errorshow]{tracefnt}
\usepackage{amsmath}
\usepackage{slashed}
\usepackage{amssymb}
\usepackage{array}
\usepackage{amsthm}
\usepackage{eucal}
\usepackage{epsf}
\usepackage{epsfig}
\usepackage{psfrag}
\usepackage{multirow}
\usepackage{wasysym} 
\usepackage{tikz-cd} 
\usepgfmodule{shapes}
\usetikzlibrary{backgrounds, arrows,calc,shapes,decorations.pathreplacing, decorations.markings, automata,positioning}

\usetikzlibrary{arrows,calc}
\usepackage{relsize}

\newcommand{\be}{\begin{equation}}
\newcommand{\ee}{\end{equation}}
\newcommand{\bea}{\begin{eqnarray}}
\newcommand{\eea}{\end{eqnarray}}
\newcommand{\ba}{\begin{array}}
\newcommand{\ea}{\end{array}}
\newcommand{\bpic}{\begin{tikzpicture}}
\newcommand{\epic}{\end{tikzpicture}}

\def\Tr{\mathrm{Tr}}
\newcommand{\tr}{\Tr\,}

\newcommand{\M}{{\mathfrak M}}

\newcommand\qt{\tilde q}

\newcommand\bt{\tilde b}

\newcommand\Qt{\tilde Q}

\def\t{\tilde}
\def\tr{\Tr}
\newcommand{\overbar}[1]{\mkern 1.5mu\overline{\mkern-1.5mu#1\mkern-1.5mu}\mkern 1.5mu}
\newcommand{\CN}{{\mathcal N}}

\newcommand{\CT}{{\mathcal T}}

\newcommand{\CW}{{\mathcal W}}

\newcommand{\cB}{\mathcal{B}}
\newcommand{\cC}{\mathcal{C}}

\newcommand{\cM}{\mathcal{M}}
\newcommand{\cN}{\mathcal{N}}

\newcommand{\cT}{\mathcal{T}}

\newcommand{\cV}{\mathcal{V}}
\newcommand{\cW}{\mathcal{W}}

\def\Bt{\tilde{B}}

\title{Monopoles and dualities in $3d$ $\cN=2$ quivers}

\author[1]{Sergio Benvenuti}
\author[2]{Ivan Garozzo}
\author[3,4]{Gabriele Lo Monaco}

\affiliation[1]{INFN Sezione di Trieste, via Bonomea 265, 34136 Trieste, Italy}
\affiliation[2]{Dipartimento di Fisica, Universit\`a di Milano-Bicocca $\&$ INFN, Sezione di Milano-Bicocca,
I-20126 Milano, Italy}
\affiliation[3]{Institut de Physique Th\'eorique, Universit\'e Paris Saclay, CEA, CNRS, \\ Orme des Merisiers, 91191 Gif-sur-Yvette CEDEX, France}
\affiliation[4]{Department of Physics, Stockholm University, AlbaNova, 10691 Stockholm, Sweden}

\emailAdd{benve79@gmail.com}
\emailAdd{i.garozzo@gmail.com}
\emailAdd{gabriele.lomonaco@ipht.fr}

\abstract{Seiberg-like dualities in $2+1$d quiver gauge theories with $4$ supercharges are investigated. We consider quivers made of various combinations of classical gauge groups $U(N)$, $Sp(N)$, $SO(N)$ and $SU(N)$. Our main focus is the mapping of the supersymmetric monopole operators across the dual theories. There is a simple general rule that encodes the mapping of the monopoles upon dualizing a single node. This rule dictates the mapping of all the monopoles which are not dressed by baryonic operators. We also study more general situations involving baryons and baryon-monopoles, focussing on three examples: $SU-Sp$, $SO-SO$ and $SO-Sp$ quivers.}

\begin{document}
\maketitle

\section{Introduction and results}
Quiver quantum field theories are gauge theories with product gauge group and matter content in rank-$2$ representations, such as bifundamental and adjoint representations. Since strings can end on two different branes, quivers are ubiquitous in string theory compactifications. For this reason, quiver theories have been studied extensively during the past three decades.

In this note we focus on $2\!+\!1$ dimensional quivers with four supercharges, that is $3d$ $\CN=2$ supersymmetry \cite{deBoer:1997ka, deBoer:1997kr, Aharony:1997bx, Aharony:1997gp}. $3d$ $\CN=2$ quivers have been recently studied, for instance, in \cite{Benini:2011mf, Amariti:2014lla, Benvenuti:2016wet, Amariti:2017gsm, Benvenuti:2017kud,Benvenuti:2017bpg, Zenkevich:2017ylb, Aprile:2018oau, Amariti:2019pky, Pasquetti:2019uop, Pasquetti:2019tix, Jain:2019lqb}. 

Theories with four supercharges, in $3\!+\!1$ or less dimensions, enjoy Seiberg dualities 
\cite{Seiberg:1994pq, Intriligator:1995id, Intriligator:1995au,Aharony:1997gp,Giveon:2008zn,Aharony:2011ci,Benini:2011mf,Aharony:2014uya}, relating two different UV theories with a single gauge group which flow to the same IR superconformal field theory. Applying Seiberg duality to a node of a quiver gauge theory, one gets a new quiver theory with the same number of nodes. 

We study examples of $3d$ $\CN=2$ quiver dualities. In each case we work out the map of chiral ring generators (in the algebraic sense) across the duality, \emph{i.e.} from one theory and a one obtained by dualizing a specific node in the quiver. It is quite easy to map \emph{mesonic operators}, which also exists in $3\!+\!1$ dimensions. In $3d$ gauge theories, however, there are also \emph{monopole operators} \cite{Borokhov:2002ib, Borokhov:2002cg}, that is local disorder operators which under duality can map to standard operators polynomial in the elementary fields . 

One of the main theme of this paper is the mapping of the monopole operators under Seiberg duality inside a quiver. In the case of linear quivers with unitary gauge groups such issue was important in a series of recent works \cite{Benvenuti:2017kud,Benvenuti:2017bpg, Zenkevich:2017ylb, Aprile:2018oau, Pasquetti:2019uop, Pasquetti:2019tix}. The process of applying a duality on a node of a quiver requires to take into account possible contact terms that may become non-trivial BF couplings when the duality is applied inside a quiver \cite{Closset:2012vg, Closset:2012vp, Amariti:2014lla} (we will be more explicit about this in the main text).

We investigate various quivers with two gauge groups, studying in each example two different duals. In the original model we take the superpotential to be vanishing. It is possible to turn on superpotential or real mass deformations, even if we do not study such deformed dualities in this paper.

The gauge groups we consider are a combination of classical groups $U(N)$, $Sp(N)$, $SO(N)$ and $SU(N)$. Although we study quivers with only two nodes, the results allow to find the general rule for the mapping of supersymmetric monopole operators under Seiberg duality. The rule is valid for quivers with an arbitrary number of gauge groups and generalizes the findings of \cite{Pasquetti:2019uop, Pasquetti:2019tix}. The quivers need not be linear, but let us state the \emph{monopole mapping rule} in the case of linear quivers, since it is simpler. 

We denote monopoles in a linear quiver with the notation $\M^{0,0,\bullet,\bullet,\ldots}$: a $0$ in the $i^{th}$ position means that there is vanishing GNO flux for the $i^{th}$ gauge group, while a $\bullet$ in the $i^{th}$ position means that there is \emph{minimal} GNO flux for the $i^{th}$ gauge group. For unitary gauge groups $U(N)$ the $\bullet$'s can be either all $+$'s or all $-$'s.

The monopoles which are chiral ring generators have minimal GNO fluxes for each node, and the non-zero fluxes are turned on in a single connected group of nodes (of arbitrary length), of the form $\M^{\ldots,0,0,\bullet,\bullet, \bullet,0,\ldots}$.
 Dualising node $i$,  the rule is as follows:
\begin{itemize}
\item a monopole with zero flux under $i-1,i,i+1$, stays the same: \newline $\qquad \M^{\ldots,0_{i-1},0_i,0_{i+1},\ldots} \to \M^{\ldots,0_{i-1},0_i,0_{i+1},\ldots}$
\item a monopole with flux under a neighbour of $i$, but not flux $i$, ``extends", that is it picks up flux under node $i$:  $\M^{\ldots,\bullet,0_i,0,\ldots} \to \M^{\ldots,\bullet,\bullet_i,0,\ldots}$,  $\M^{\ldots,0,0_i,\bullet,\ldots} \to \M^{\ldots,0,\bullet_i,\bullet,\ldots}$.
\item a monopole with flux under $i$ but not $i+1$ or $i-1$, $\M^{\ldots,0,\bullet_i,0,\ldots}$, maps to a gauge singlet operator flipping the monopole $\M^{\ldots,0,\bullet_i,0,\ldots}$.
\item a monopole with flux under $i$ and $i+1$ or  $i$ and $i-1$\footnote{Notice that we do not consider $\M^{\ldots,\bullet,0_i,\bullet,\ldots}$. This operator is not a chiral ring generator, its mapping is obtained from the mapping of the two monopoles $\M^{\ldots,\bullet,0_i,0,\ldots}$ and $\M^{\ldots,0,0_i,\bullet,\ldots}$.} ``shortens", that is it loses the flux under node $i$:  $\M^{\ldots,\bullet,\bullet_i,0,\ldots} \to \M^{\ldots,\bullet,0_i,0,\ldots}$,  $\M^{\ldots,0,\bullet_i,\bullet,\ldots} \to \M^{\ldots,0,0_i,\bullet,\ldots}$.
\item a monopole with flux under $i-1,i,i+1$ stays the same: $\M^{\ldots,\bullet,\bullet_i,\bullet,\ldots} \to \M^{\ldots,\bullet,\bullet_i,\bullet,\ldots}$.
\end{itemize}

From the mapping of these basic monopoles,  the mapping of generic dressed monopoles follows.

Let us emphasize that the rule stated above is not enough to deal with  \emph{baryonic operators} and \emph{baryon-monopoles}, which are present when there is a gauge group defined by the determinant $=1$ condition ($SU(N)$ or $SO(N)$), or when $U(N)$ is present together with $O$, $Sp$, $SU$, $SO$.\footnote{Theories with alternating $Sp/(S)O$ groups appears naturally in the context of $3d$ $\mathcal N=4$ theories via their realization as gauge theories on the $D3$ branes on top of orientifold planes. A large class of such theories is realized in the context of $S$-duality walls in 
$\mathcal N=4$ SYM with real gauge groups \cite{Gaiotto:2008ak}. Various properties of such $3d$ theories with 8 supercharges have been widely discussed \cite{Cremonesi:2014kwa, Cremonesi:2014vla, Cremonesi:2014uva, Garozzo:2018kra}. Applications to theories with four supercharges are studied in \cite{Giacomelli:2017vgk}.}
We study various examples of quivers with baryons and baryon-monopoles, with two gauge groups. In these specific examples we are able to find the full mapping of the chiral ring generators. One interesting observation (Section \ref{SOSO}) is that in the case of $SO-SO$ quivers a monopole operators with three magnetic fluxes  (two non-zero fluxes in one $SO$ node and one non-zero flux in the other $SO$ node) turned on is a chiral ring generator. This is to be contrasted with the $SO$ gauge theory, where the chiral ring generators have at most one magnetic flux turned on.

\vspace{0.2cm}

%A more general study of quivers with baryons and baryon-monopoles goes beyond the purpose of this paper.

We will present an application of the results in this note in two companion papers \cite{BGLMseqdec:2020, BGLMseqdec:2020b}. In \cite{BGLMseqdec:2020} we study the duals of a theory with a single gauge group and matter in the a rank-$2$ representation, $U(N)$ with a single adjoint field and flavors, or $Sp(N)$ with a single antisymmetric field and flavors. We produce a dual which is a quiver with $N$ gauge nodes, obtained by \emph{sequentially deconfining} the rank-$2$ field. The process requires to use many times a Seiberg-like duality inside a quiver, and at each step, in order to control the superpotential, it is crucial to have the mapping of all the chiral ring operators, including the monopoles.

\vspace{0.6cm}

The paper is organized as follows.

\vspace{0.2cm}

In section \ref{typeI} we consider quivers with $U(N)$ or $Sp(N)$ gauge groups (so there are no baryonic operators, and no monopoles dressed by baryonic operators). In each case we study in detail a two-node quiver with flavors: we produce two different dual theories, obtained by dualizing the left or the right node. We tried to make each subsection readable independently from the other.

\vspace{0.2cm}

In section \ref{typeII} we study quivers with baryons and baryon-monopoles, including $SO$ and $SU$ gauge groups. We work out three examples, with gauge groups $SO (N_1) \times Sp(N_2)$, $SO (N_1) \times SO(N_2)$ and $Sp (N_1) \times SU(N_2)$. In each example we are able to find the mapping of all the chiral ring generators. 

\vspace{0.2cm}

The main tool we employ to perform our analysis is the $3d$ supersymmetric index, that can be computed as the partition function on $S^2 \times \mathbb R$ \cite{Bhattacharya:2008zy, Bhattacharya:2008bja, Kim:2009wb, Imamura:2011su, Kapustin:2011jm, Dimofte:2011py, Aharony:2013dha, Aharony:2013kma}, whose fundamental definitions and properties we review in appendix \ref{appA}.

%%%%%%%%%%%%%%%%%%%%%%%%%%%%%%%%%%%%%%%%%%%%%%%%%%%%%%%%%%%%%%%
%%%%%%%%%%%%%%%%%%%%%%%%%%%%%%%%%%%%%%%%%%%%%%%%%%%%%%%%%%%%%%%

\section{Duality in quivers without baryons}\label{typeI}
In this section we study quivers with gauge groups $U(N)$ and $Sp(N)$. With this choice of gauge groups, there are no baryonic operators, and no monopoles dressed by baryonic operators (we consider examples with special gauge groups, $SU(N)$ or $SO(N)$, and hence baryon and baryon-monopoles operators, in section \ref{typeII}).

In each case we study in detail a two node quiver with flavors: we produce two different dual theories, obtained by dualizing the left or the right node. We discuss in great detail the global symmetries of the supersymmetric monopole operators which are generators of the chiral ring and we exhibit the map of the full set of chiral generators. This is enough to find the general rule for the mapping of the chiral ring operators in arbitrary quivers, which is the main goal of this paper.

\subsection{Unitary gauge groups}\label{UU}

\be
\scalebox{1}{
\begin{tikzpicture}[baseline]
\tikzstyle{every node}=[font=\footnotesize]
\node[draw=none] (node1) at (0,0) {$U(N_1)$};
\node[draw=none] (node2) at (2.5,0) {$U(N_2)$};
\node[draw, rectangle] (sqnode) at (5,0) {$F$};
%\draw[black,solid] (node) edge [out=45,in=135,loop,looseness=5]node[above]{$\Phi$} (node) ;
\draw[black,solid,->] (node1) edge [out=25,in=155,loop,looseness=1]node[above]{$B$} (node2) ;
\draw[black,solid,<-] (node1) edge [out=-25,in=-155,loop,looseness=1]node[below]{$\t B$} (node2) ;
\draw[black,solid,->] (node2) edge [out=25,in=155,loop,looseness=1]node[above]{$Q$} (sqnode) ;
\draw[black,solid,<-] (node2) edge [out=-25,in=-155,loop,looseness=1]node[below]{$\t Q$} (sqnode) ;
\node[draw=none] at (2.5,-1.5) {$\mathcal W=0$};
\node[draw=none] at (-1.5,0) {$\mathcal T_A:$};
\end{tikzpicture}}
\ee
We assume $N_2 \ge N_1$ otherwise the strongly coupled gauge dynamics of the left node breaks supersymmetry (via a runaway potential). The global symmetry group is $SU(F) \times SU(F) \times U(1)_{B} \times U(1)_{T_1} \times U(1)_{Q}  \times U(1)_{T_2}$, where $U(1)_{B}$ and $U(1)_{Q}$ are the axial symmetries acting on $B, \, \t B$ and $Q, \t Q$ and 
$U(1)_{T_1}$ and $U(1)_{T_2}$ are the topological symmetries for $U(N_1)$ and $U(N_2)$ respectively.

The $R$-charge of monopole operators $\mathfrak M^{\vec m, \vec n}$ reads
\be
\begin{split}
	R[\mathfrak M_A^{\vec m, \vec n}]&=2\times\frac{1}{2}(1-R_B)\sum_{a=1}^{N_1}\sum_{b=1}^{N_2} 
	|m_a-n_b| + 2\times\frac{1}{2}F(1-R_Q)\sum_{b=1}^{N_2} |n_b|+ \\ 
	&-\sum_{a_1<a_2}|m_{a_1}-m_{a_2}|-\sum_{b_1<b_2}|n_{b_1}-n_{b_2}|,
\end{split}
\ee
Specialising this formula for the monopoles with minimal GNO flux $\mathfrak M^{\pm, 0}, \; \mathfrak M^{0,\pm}, \;
\mathfrak M^{\pm, \pm}$ we get
\begin{align}
	&R[\mathfrak M_A^{\pm, 0}]=N_2 (1-R_B) - (N_1-1), \\
	&R[\mathfrak M_A^{0, \pm}]=N_1(1-R_B)+F (1-R_Q) - (N_2-1), \\
	&R[\mathfrak M_A^{\pm, \pm}]=(N_1+N_2-2) (1-R_B)+F (1-R_Q) - (N_1-1) - (N_2-1),
\end{align}
The chiral ring generators include both mesons and monopole operators. The former ones are represented by dressed mesons of the form $\tr (Q_i (B \t B)^J Q_j)$ for $J=0,\dots, N_1$ and 
$\tr \left((B \t B)^J\right)$ with the same range for $J$. The monopole generating the chiral ring are $\mathfrak M_A^{0,\pm}, \,\mathfrak M_A^{\pm,0}\, \mathfrak M_A^{\pm,\pm}$. In the following we will discuss the two Aharony dual frames and discuss the map of chiral ring generators.

\subsubsection*{First dual}
Let us apply Aharony duality to the $U(N_1)$ node, we get: 
\be
\scalebox{1}{
\begin{tikzpicture}[baseline]
\tikzstyle{every node}=[font=\footnotesize]
\node[draw=none] (node1) at (0,0) {$U(N_2-N_1)$};
\node[draw=none] (node2) at (3,0) {$U(N_2)$};
\node[draw, rectangle] (sqnode) at (6,0) {$F$};
\draw[black,solid,->] (node1) edge [out=25,in=155,loop,looseness=1]node[above]{$b$} (node2) ;
\draw[black,solid,<-] (node1) edge [out=-25,in=-155,loop,looseness=1]node[below]{$\t b$} (node2) ;
\draw[black,solid] (node2) edge [out=45,in=135,loop,looseness=5]node[above]{$\phi_b$} (node2) ;
\draw[black,solid,->] (node2) edge [out=25,in=155,loop,looseness=1]node[above]{$Q$} (sqnode) ;
\draw[black,solid,<-] (node2) edge [out=-25,in=-155,loop,looseness=1]node[below]{$\t Q$} (sqnode) ;
\node[draw=none] at (3,-1.5) {$\mathcal W=\sigma_B^{\pm} \mathfrak M_B^{\pm, 0} +
\Tr(\t b \phi_b b)$};
\node[draw=none] at (-1.5,0) {$\mathcal T_B:$};
\end{tikzpicture}}
\ee
where $\phi_b$ is a traceful adjoint field. 
The $R$-charge of monopole operators $\mathfrak M_B^{\vec m, \vec n}$ reads
\be\label{UUMonTB}
\begin{split}
	R[\mathfrak M_B^{\vec m, \vec n}]&=2\times\frac{1}{2}(1-R_b)
	\sum_{a=1}^{N_2-N_1}\sum_{b=1}^{N_2} 
	|m_a-n_b| + 2\times\frac{1}{2}F(1-R_Q)\sum_{b=1}^{N_2} |n_b|+ \\ 
	&+2\times\frac{1}{2}(1-R_{\phi_b})\sum_{a_1<a_2}|m_{a_1}-m_{a_2}|-\sum_{a_1<a_2}|m_{a_1}-m_{a_2}|-\sum_{b_1<b_2}|n_{b_1}-n_{b_2}|.
\end{split}
\ee
The basic monopole operators have the following $R$-charges
\begin{align}
	&R[\mathfrak M_B^{\pm, 0}]=N_2 (1-R_b) - (N_2-N_1-1), \\
	&R[\mathfrak M_B^{0, \pm}]=(N_2-N_1)(1-R_b)+F (1-R_q)+(N_2-1)(1-R_{\phi_b}) - (N_2-1), \end{align}
\be
\begin{split}
	R[\mathfrak M_B^{\pm, \pm}]&=(2N_2-N_1-2) (1-R_b)+F (1-R_q)+(N_2-1)(1-R_{\phi_b})+ \\
	&- (N_2-N_1-1) - (N_2-1).
\end{split}
\ee
The $R$-charges of fundamental fields can be mapped to the ones of theory $\mathcal T_A$ using that 
\be
	B \t B \leftrightarrow \phi_b, \qquad \Tr(Q_i \t Q^j) \leftrightarrow \Tr(q_i \t q^j),
	\qquad 
\ee
and the superpotential term $\Tr(\t b \phi_b b)$, implying that 
\be\label{mapUU1}
	R_b=1-R_B, \qquad R_{\phi_b}=2 R_B, \qquad R_q=R_Q.
\ee
The mesonic part of the chiral ring is generated by dressed operators $\tr (q_i \phi_b^J \t q^j)$ and 
$\tr(\phi_b^J)$, for $J=0, \dots, N_2$.
As already discussed in the case of the triality for $Sp(N_1) \times Sp(N_2)$ gauge theory, the  chiral ring generators include dressed monopoles. In detail, we have the singlets $\sigma_B^{\pm}$ flipping the monopoles $\mathfrak M_B^{\pm, 0}$ and the monopoles with flux on both gauge nodes $\mathfrak M_B^{\pm, \pm}$. The monopoles $\mathfrak M_B^{0, \pm}$ are the ones that can be dressed with the adjoint field $\phi_B$, and we get the tower $\{(\mathfrak M_B^{0, \pm})_{\phi_B^J}\}$ with $J=0, \dots, N_2$. 
%we can see, by using \eqref{mapUU1} that monopoles with fluxes on both nodes in theory $\mathcal T_A$ map to monopoles with flux only on one node dressed with adjoint field. 
%In detail, one has the following map
%\be
%\begin{tabular}{c c c}
%$\mathcal T_A$ & & $\mathcal T_B$ \\
%\hline
%$\Tr(Q_i Q^j)$ & & $\Tr(q_i q^j)$ \\
%$\Tr\left((B \t B)^J\right)$ & & $\Tr(\phi_b^J)$ \\
%$\Tr(Q_i (B \t B)^J \t Q^j)$ & & $\Tr(q_i \phi_b^{J} \t q^j)$ \\
%%$\Tr\left((B B)^J\right)$ & & $\Tr(\phi_c^J)$ \\
%$\mathfrak M_A^{\pm, 0}$ & & $\sigma_B^\pm$ \\
%$\mathfrak M_A^{0, \pm}$ & & $\mathfrak M_B^{\pm, \pm}$ \\
%$\{(\mathfrak M_A^{\pm, \pm})_{(B \t B)^J} \}$ & & $\{(\mathfrak M_B^{0, \pm})_{\phi_b^J} \}$ \\
%\end{tabular}
%\ee
Observe that in theory $\mathcal T_B$ there is no dressed monopole of the form 
$\{(\mathfrak M_B^{\pm, \pm})_{(b \t b)^J}\}$ since the $F$-term equations set to zero the matrix $bb$ on the chiral ring.

\subsubsection*{Second dual}
\be
\scalebox{1}{
\begin{tikzpicture}[baseline]
\tikzstyle{every node}=[font=\footnotesize]
\node[draw=none] (node1) at (0,0) {$U(N_1)$};
\node[draw=none] (node2) at (4,0) {$U(N_1-N_2+F)$};
\node[draw=none] (node21) at (3.8,0.4) {};
\node[draw, rectangle] (sqnode) at (2,3) {$F$};
\draw[black,solid,->] (node1) edge [out=25,in=155,loop,looseness=0.6]node[below]{$b$} (node2) ;
\draw[black,solid,<-] (node1) edge [out=-25,in=-155,loop,looseness=0.6]node[below]{$\t b$} (node2) ;
\draw[black,solid] (sqnode) edge [out=45,in=135,loop,looseness=5]node[above]{$M$} (sqnode) ;
\draw[black,solid] (node1) edge [out=-135,in=135,loop,looseness=5]node[left]{$\phi_c$} (node1) ;
\draw[black,solid,->] (node1) edge [out=75,in=215,loop,looseness=0.6]node[above]{$p$} (sqnode) ;
\draw[black,solid,<-] (node1) edge [out=25,in=265,loop,looseness=0.6]node[below]{$\t p$} (sqnode) ;
\draw[black,solid,->] (node21) edge [out=165,in=-90,loop,looseness=0.6]node[below]{$q$} (sqnode) ;
\draw[black,solid,<-] (node21) edge [out=75,in=-30,loop,looseness=0.6]node[right]{$\t q$} (sqnode) ;
\node[draw=none] at (2,-1.5) {$\mathcal W=\sigma_C^{\pm} \mathfrak M_C^{0,\pm} +
\Tr(\t b \phi_c b) + \Tr(b p q)+\Tr(\t b\, \t p\, \t q) + \Tr(q M \t q)$};
\node[draw=none] at (-3,0) {$\mathcal T_C:$};
\end{tikzpicture}}
\ee
The formula for the $R$-charge of a monopole with general fluxes is completely analogous to \eqref{UUMonTB}, hence we do not repeat it here. We write down the $R$-charges of monopoles with minimal GNO flux
\be
	R[\mathfrak M_C^{\pm, 0}]=(N_1-N_2+F)(1-R_b) + (N_1-1)(1-R_{\phi_c}) + F(1-R_p) 
	-(N_1-1),
\ee
\be
	R[\mathfrak M_C^{0, \pm}]=N_1(1-R_b) + F(1-R_q) -(N_1-N_2+F-1),
\ee
\be
\begin{split}
	R[\mathfrak M_C^{\pm, \pm}]&=(2N_1-N_2+F-2)(1-R_b) +(N_1-1)(1-R_{\phi_c})+ F(1-R_q)+\	\\
	&+F(1-R_p)-(N_1-1)-(N_1-N_2+F-1).
\end{split}
\ee
The $R$-charges of fundamental fields can be mapped to the ones of theory $\mathcal T_A$ using the map of mesons to singlets implied by Aharony duality 
\be
	\tr(Q_i \t Q^j) \leftrightarrow M^i_j, \qquad \Tr(B \t B) \leftrightarrow \Tr(\phi_c), 
\ee
combining this piece of information to the superpotential terms in theory $\mathcal T_C$ we find 
\be
	R_b=1-R_B, \qquad R_q=1-R_Q, \qquad R_p=R_B+R_Q, \qquad R_{\phi_c}=2 R_B. 
\ee
The chiral ring generators include the singlets $M^i_j$ and $\sigma_B^\pm$, the mesonic-like operators 
$\tr(\phi_c^J)$ and $\tr(p_i \phi_c^J \t p^j)$ for $J=0, \dots, N_1$. The monopole operators generating the chiral ring are similar to the ones discussed for $\mathcal T_B$: we have the ones with flux on the gauge nodes $\mathfrak M_C^{\pm, \pm}$ and the ones dressed with the adjoint field $\phi_C$: $\{(\mathfrak M_C^{\pm, 0})_{\phi_c^J} \}$.

\subsubsection*{Operator map}
In order to discuss the map of chiral ring generators across the triality it is useful to write down the map of the $R$-charges of the fundamental fields
\begin{itemize}
\item{$\mathcal T_B \to \mathcal T_A$:
\be	
	R_b=1-r_B, \qquad R_{\phi_b}=2R_B.
\ee
}
\item{$\mathcal T_C \to \mathcal T_A$:
\be	
	R_b=1-R_B, \qquad R_q=1-R_Q, \qquad R_p=R_B+R_Q, \qquad R_{\phi_c}=2 R_B.
\ee
}
\end{itemize}
Using the $R$-charge map we have just summarised and the representation of the global symmetry group under which the various operators transform we can study how the chiral ring generators map across duality. The complete mapping is the following one
\be
\begin{tabular}{c c c c c}
$\mathcal T_A$ & & $\mathcal T_B$ & & $\mathcal T_C$ \\
\hline
$\Tr(Q_i Q^j)$ & & $\Tr(q_i q^j)$ & & $M_i^j$ \\
$\Tr\left((B \t B)^J\right)$  & & $\Tr(\phi_b^J)$& & $\Tr(\phi_c^{J})$ \\
$\Tr(Q_i (B \t B)^J \t Q^j)$  & & $\Tr(q_i \phi_b^{J} \t q^j)$& & $\Tr(p_i \phi_c^{J-1} \t p^j)$ \\
%$\Tr\left((B B)^J\right)$ & & $\Tr(\phi_c^J)$ \\
$\mathfrak M_A^{\pm, 0}$  & & $\sigma_B^\pm$& & $\mathfrak M_C^{\pm, \pm}$ \\
$\mathfrak M_A^{0, \pm}$  & & $\mathfrak M_B^{\pm, \pm}$& & $\sigma_C^\pm$ \\
$\{(\mathfrak M_A^{\pm, \pm})_{(B \t B)^J} \}$ & & $\{(\mathfrak M_B^{0, \pm})_{\phi_b^J} \}$ & & $\{(\mathfrak M_C^{\pm, 0})_{\phi_c^J} \}$ \\
\end{tabular}
\ee
In the next section we will show the supersymmetric index for the triality just discussed with a particular emphasis on the chiral ring generators.

\subsubsection*{Supersymmetric index}
In this last section we compute the supersymmetric index for the three dual frames $\mathcal T_A$, $\mathcal T_B$ and $\mathcal T_C$ for $N_1=1, \, N_2=2, \, F=2$ and with the choice of $R$-charges given by $R_B=1/5, \, R_Q=3/8$:
\be\label{IndSpSp}
\begin{split}
		\mathcal I&=1+{\color{blue}{x^{2/5} y_B^2}}+{\color{blue}{4 x^{3/4} y_Q^2}}+x^{4/5} y_B^4+x^{21/20} \left({\color{blue}{\frac{\omega _1 \omega _2}{y_B y_Q^2}+}}{\color{blue}{\frac{\omega _2}{y_B y_Q^2}+\frac{1}{y_B y_Q^2 \omega _1 \omega _2}}}{\color{blue}{+\frac{1}{y_B y_Q^2 \omega _2}}}\right)+\\
	&+{\color{blue}{8 x^{23/20} y_B^2 y_Q^2}}+x^{6/5} y_B^6+x^{29/20} \left({\color{blue}{\frac{\omega _1 \omega _2 y_B}{y_Q^2}}}+\frac{\omega _2 y_B}{y_Q^2}+{\color{blue}{\frac{y_B}{y_Q^2 \omega _1 \omega _2}}}+\frac{y_B}{y_Q^2 \omega _2}\right)+10 x^{3/2} y_Q^4+\\
	&+8 x^{31/20} y_B^4 y_Q^2+x^{8/5} \left({\color{blue}{\frac{\omega _1}{y_B^2}+\frac{1}{\omega _1 y_B^2}}}+y_B^8\right)+x^{9/5} \left(\frac{4 \omega _1 \omega _2}{y_B}+\frac{4 \omega _2}{y_B}+\frac{4}{y_B \omega _2}+\frac{4}{y_B \omega _1 \omega _2}\right)+\\
	&+x^{37/20} \left(\frac{\omega _1 \omega _2 y_B^3}{y_Q^2}+\frac{\omega _2 y_B^3}{y_Q^2}+\frac{y_B^3}{y_Q^2 \omega _1 \omega _2}+\frac{y_B^3}{y_Q^2 \omega _2}\right)+25 x^{19/10} y_B^2 y_Q^4+8 x^{39/20} y_B^6 y_Q^2+x^2 \left(y_B^{10}-10\right)+ \dots
	\end{split}
\ee

where $y_B$ and $y_Q$ are the fugacities for the two axial symmetries. The charges of the fundamental fields under these symmetries are assigned as follows:
\be
\begin{tabular}{c c c}
$$ & $U(1)_{B}$ & $U(1)_{Q}$ \\
\hline
$B, \Bt$ & $1$ & $0$ \\ 
$Q, \Qt$ & $0$ & $1$ \\ 
\hline
$b, \bt$ & $-1$ & $0$ \\ 
$q, \qt$ & $0$ & $1$ \\ 
\hline
$b, \bt$ & $-1$ & $0$ \\ 
$q, \qt$ & $0$ & $-1$ \\ 
\end{tabular}
\ee
The topological symmetries map is straightforward in going from $\mathcal T_A$ to $\mathcal T_B$, while in going from $\mathcal T_A$ to $\mathcal T_C$ they mix non-trivially in a way that can be summarised in the following fugacity map
\be
	\omega_1^{\mathcal T_C} \to \frac{1}{\omega_1^{\mathcal T_A}}, \quad 
	\omega_2^{\mathcal T_C} \to \omega_1^{\mathcal T_A} \omega_2^{\mathcal T_A},
\ee
where the label 1 and 2 refers to the $U(N_1)$ and $U(N_2)$ for $\mathcal T_A$ and 
$U(N_1)$ and $U(N_1-N_2+F)$ for $\mathcal T_C$.
In \eqref{IndSpSp} we highlighted in blue the various chiral ring generators already discussed in the previous sections. For definiteness, let us take the generators in the frame $\mathcal T_A$ and identify them with the various terms we find in the supersymmetric index
\begin{align}
	\Tr(Q_i\t Q^j) \;&\leftrightarrow\; {\color{blue}{4 x^{3/4} y_Q^2}} \\
	\Tr(B\t B) \;&\leftrightarrow \;{\color{blue}{x^{2/5} y_B^2}} \\
	\Tr(Q_i(B\t B)\t Q^j) \;&\leftrightarrow \;{\color{blue}{8 x^{23/20} y_B^2 y_Q^2}} \\
	\mathfrak M_A^{\pm, 0} \;&\leftrightarrow \; {\color{blue}x^{8/5} \left(\frac{\omega _1}{y_B^2}+\frac{1}{y_B^2 \omega _1}\right)} \\
	\mathfrak M_A^{0, \pm} \;&\leftrightarrow \;  {\color{blue}{x^{21/20} \left(\frac{\omega _2}{y_B y_Q^2}+\frac{1}{y_B y_Q^2 \omega _2}\right)}} \\
	\mathfrak M_A^{\pm, \pm} \;&\leftrightarrow \; {\color{blue}{x^{21/20} \left(\frac{\omega _1 \omega _2}{y_B y_Q^2}+\frac{1}{y_B y_Q^2 \omega _1 \omega _2}\right)}} 
	\\
	\{(\mathfrak M_A^{\pm, \pm})_{BB} \}\;&\leftrightarrow
	{\color{blue}{x^{29/20} \left(\frac{\omega _1 \omega _2 y_B}{y_Q^2}+\frac{y_B}{y_Q^2 \omega _1 \omega _2}\right)}}.
\end{align}

%%%%%%%%%%%%%%%%%%%%%%%%%%%%%%%%%%%%%%%%%%%%%%%%%%%%%%%%%%%%%%%
\subsection{Symplectic gauge groups}\label{SPSP}
Consider a theory with two gauge groups and flavours
\be
\scalebox{1}{
\begin{tikzpicture}[baseline]
\tikzstyle{every node}=[font=\footnotesize]
%\node[draw, rectangle] (sqnodeL) at (-4,0) {$2$};
\node[draw=none] (node1) at (0,0) {$Sp(N_1)$};
\node[draw=none] (node2) at (3,0) {$Sp(N_2)$};
\node[draw, rectangle] (sqnode) at (6,0) {$2F$};
\draw[black,solid] (node1) edge node[above]{$B$} (node2) ;
\draw[black,solid] (node2) edge node[above]{$Q$} (sqnode) ;
%\draw[black,solid,->] (node2) edge [out=25,in=155,loop,looseness=1]node[above]{$F$} (sqnodeR) ;
\node[draw=none] at (3,-1) {$\mathcal W=0$};
\node[draw=none] at (-2,0) {$\mathcal T_{A}:$};
%node[above]{\blue $X$} node[below, yshift=0cm] {$\red F_X$} node {$ \red \times$}  (node);
\end{tikzpicture}}
\ee
We assume $N_2 \ge N_1 + 1$, otherwise the strongly coupled dynamics of the left node breaks 
supersymmetry via a runaway superpotential \cite{Affleck:1982as}. It is possible to find two dual descriptions of this theory by applying Aharony duality \cite{Aharony:1997gp} to each gauge node. Similar techniques of studying quiver theories through node by node dualisation using Aharony and one-monopole duality \cite{Benini:2017dud} have been applied in order to get the dual of $U(N)$ with one flavour and adjoint \cite{Pasquetti:2019uop} and the generalisation to $k+1$ flavours \cite{Pasquetti:2019tix}.

%under which the fundamental fields are charged as
%\be
%\begin{tabular}{c c c c}
% & $SU(2F)$ & $U(1)_{A_1}$ & $U(1)_{A_2}$ \\
%\hline
%$B$ & $\mathbf{1}$ & $1$ & $0$ \\
%$Q$ & $\mathbf{2F}$ & $0$ & $1$ \\
%$\mathfrak M_A^{\cdot, 0}$ & $\mathbf{1}$ & $-2N_2$ & $0$ \\
%$\mathfrak M_A^{0,\cdot}$ & $\mathbf{1}$ & $-2N_1$ & $-2F$ \\
%$\mathfrak M_A^{\cdot, \cdot}$ & $\mathbf{1}$ & $-2(N_1+N_2-1)$ & $-2F$ \\
%\end{tabular}
%\ee
%where we used that the charge of the monopole $\mathfrak M^{\vec m, \vec n}$ under the axial symmetries $U(1)_b$ and $U(1)_f$ are
%\begin{align}
%	&A_1[\mathfrak M^{\vec m, \vec n}]=-\frac{1}{2}A_1[B]\sum_{a=1}^{N_1}
%	\sum_{b=1}^{N_2}\sum_{\sigma, \gamma=0,1}
%	|(-1)^\sigma m_a + (-1)^\gamma n_b|\\
%	&A_2[\mathfrak M^{\vec m, \vec n}]=-\frac{1}{2}A_2[Q]\sum_{b=1}^{N_2}\sum_{\gamma=0,1}
%	|(-1)^\gamma n_b|
%\end{align}
The goal of our analysis is to study the map of chiral ring generators across duality, with particular focus on the monopole operators, including the dressed ones. For theories with $\mathcal N=4$ supersymmetry this problem has been beautifully addressed in \cite{Cremonesi:2013lqa} using the Hilbert series \cite{Benvenuti:2006qr}, while in the case of $\mathcal N=2$ theories the same tool has been applied to vector-like theories \cite{Cremonesi:2015dja} and to CS theories \cite{Cremonesi:2016nbo} to discuss the structure of the moduli space of vacua.

The global symmetry group of the theory is $SU(2F) \times U(1)_B \times U(1)_Q$, where the label in the Abelian symmetries denote on which chirals it is acting on.
In order to study the chiral ring generators it is useful to write down the $R$-charge of a monopole operator with general magnetic fluxes. Denoting the most general monopole for 
$\mathcal T_A$ as $\mathfrak M^{\vec m, \vec n}$, its $R$-charge is as follows
\be \label{monTA}
\begin{split}
	R[\mathfrak M_A^{\vec m, \vec n}]&=\frac{1}{2} (1-R_B) \sum_{a=1}^{N_1}\sum_{b=1}^{N_2}
	\sum_{\sigma, \gamma={0,1}} |(-1)^\sigma m_a+(-1)^\gamma n_b|+ \\
	&+\frac{1}{2} (2F)(1-R_Q)\sum_{b=1}^{N_2}\sum_{\sigma=0,1} |(-1)^\sigma m_b |+ \\
	&-\sum_{a_1<a_2}^{N_1} \bigl(|m_{a_1}-m_{a_2}|+|m_{a_1}+m_{a_2}|\bigr) -\sum_{a=1}^{N_1} |2m_a| \\
	&-\sum_{b_1<b_2}^{N_2} \bigl(|n_{b_1}-n_{b_2}|+|n_{b_1}+n_{b_2}|\bigr)
	-\sum_{b=1}^{N_2} |2n_b|,
\end{split}
\ee
where the various contributions correspond respectively to the bifundamental $B$, the fundamentals $Q$ and the last two lines are the gauginos for the two gauge groups. In this formula $R_Q$ and $R_B$ gives the mixing coefficients of the UV $R$-symmetry with the axial symmetries. The coefficients of $R_Q$ and $R_B$ gives the charge under the corresponding axial symmetry of the monopole under study. In particular, we may specialise \eqref{monTA} to the monopoles with minimal flux \footnote{For an $Sp(N)$ theory the monopole $\mathfrak M^{\vec m}$ with minimal flux has $\vec m=(1, 0^{N-1})$.}:
\begin{align}
	&R[\mathfrak M_A^{\cdot, 0}] = 2N_2(1-R_B) - 2N_1, \\
	&R[\mathfrak M_A^{ 0, \cdot}] = 2N_1(1-R_B)+2F(1-R_Q) - 2N_2, \\
	&R[\mathfrak M_A^{\cdot, \cdot}] = 2(N_1+N_2-1)(1-R_B)+2F(1-R_Q)- 2N_1 - 2N_2.
\end{align}
%Observe that taking the derivatives of the monopole $R$-charge with respect to $r_B$ and $r_Q$  we find the charges of the monopoles under the axial symmetries. For instance, for the 
%$\mathfrak M_A^{\cdot, \cdot}$ monopole we get
%\begin{align}
%	&\partial_{r_B}R[\mathfrak M_A^{\cdot, \cdot}]=-2(N_1+N_2-1), \\
%	&\partial_{r_Q}R[\mathfrak M_A^{\cdot, \cdot}]=-2F.
%\end{align}
Let us discuss the chiral ring generators of this theory. The mesonic generators comprise $\Tr(Q_i (BB)^J Q_j)$, $\Tr\left((BB)^J\right)$ for $J=0, \dots, N_1$. There are three types monopole operators, depending on the fluxes under $Sp(N_1)$ and $Sp(N_2)$: $\mathfrak M_A^{\bullet, 0}$, $\mathfrak M_A^{0, \bullet}$ and $\mathfrak M_A^{\bullet, \bullet}$. These are the bare monopole operators, however in general there are also dressed monopoles which enters 
the chiral ring as generators. It is well known that, in the context of theories with $Sp(N)$ gauge theories with antisymmetric field $\phi$ and fundamental chirals, the chiral ring is partly generated by $\mathfrak M_{\phi^J}$ with $J=0, \dots, N$ \cite{Benvenuti:2018bav}. 
For theories with more than one gauge node there are new monopoles which enter the chiral ring. In the case of two-node quiver these are monopoles with flux on both nodes 
$\mathfrak M_A^{\bullet, \bullet}$ dressed with powers of the bifundamental chirals $(BB)^J$.
Thus, the monopoles generating the chiral ring are 
\be
	\mathfrak M_A^{\bullet, 0}, \qquad \mathfrak M_A^{0, \bullet}, \qquad 
	\{(\mathfrak M_A^{\bullet, \bullet})_{(BB)^J}\},
\ee
for $J=0, \dots, N_1$.
%It is worth to comment on the fluxes for a monopole operator in a $USp(2N)$ gauge theory. It is understood that most of the Coulomb branch is lifted due to non-perturbative corrections. The part which is not lifted is a one-dimensional lattice of magnetic charges
%\be
%	\Gamma_{USp(2N)}=\{m=(m_1, \underbrace{0,\dots, 0}_{N-1}) \, | \, m_1\ge 0\}.
%\ee
%In particular, with $\mathfrak M^{\bullet}$ we denote a monopole with minimal flux $m_1=1$.

\subsubsection*{First dual}
Applying Aharony duality to the $Sp(N_1)$ node in $\mathcal T_A$ to get theory $\mathcal T_B$ we get a dual of the quiver 
\be
\scalebox{1}{
\begin{tikzpicture}[baseline]
\tikzstyle{every node}=[font=\footnotesize]
%\node[draw, rectangle] (sqnodeL) at (-4,0) {$2$};
\node[draw=none] (node1) at (0,0) {$Sp(N_2-N_1-1)$};
\node[draw=none] (node2) at (4,0) {$Sp(N_2)$};
\node[draw, rectangle] (sqnode) at (7,0) {$2F$};
\draw[black,solid] (node1) edge node[above]{$b$} (node2) ;
\draw[black,solid] (node2) edge node[above]{$Q$} (sqnode) ;
\draw[black,solid] (node2) edge [out=45,in=135,loop,looseness=5]node[above]{$\phi_b$} (node2) ;
%\draw[black,solid,->] (node2) edge [out=25,in=155,loop,looseness=1]node[above]{$F$} (sqnodeR) ;
\node[draw=none] at (4,-1) {$\mathcal W=\sigma_B \, \mathfrak M_B^{\bullet, 0} + \Tr(b \phi_b b)$};
\node[draw=none] at (-3,0) {$\mathcal T_{B}:$};
\end{tikzpicture}}
\ee
where $\phi_b$ is a traceful antisymmetric field generated by the Aharony duality.  The $R$-charge of the monopole $\mathfrak M_{B}^{\vec m, \vec n}$ reads
\be\label{RchMonB}
\begin{split}
	R[\mathfrak M_B^{\vec m, \vec n}]&=\frac{1}{2} (1-R_{b}) \sum_{a=1}^{N_2-N_1-1}\sum_{b=1}^{N_2}
	\sum_{\sigma, \gamma={0,1}} |(-1)^\sigma m_a+(-1)^\gamma n_b|+ \\
	&+\frac{1}{2} (2F)(1-R_{Q})\sum_{b=1}^{N_2}\sum_{\sigma=0,1} |(-1)^\sigma m_b |+
	\frac{1}{2} (1-R_{\phi_b}) \sum_{b_1<b_2}^{N_2} \sum_{\sigma, \gamma=0,1}
	|(-1)^\sigma n_{b_1} + (-1)^\gamma n_{b_2}| \\
	&-\sum_{a_1<a_2}^{N_2-N_1-1} \bigl(|m_{a_1}-m_{a_2}|+|m_{a_1}+m_{a_2}|\bigr) -
	\sum_{a=1}^{N_2-N_1-1} |2m_a| \\
	&-\sum_{b_1<b_2}^{N_2} \bigl(|n_{b_1}-n_{b_2}|+|n_{b_1}+n_{b_2}|\bigr)
	-\sum_{b=1}^{N_2} |2n_b|.
\end{split}
\ee
The dimensions of monopoles with minimal fluxes are given by
\begin{align}
	&R[\mathfrak M_B^{\bullet, 0}] = 2N_2(1-R_{b}) - 2(N_2-N_1-1), \\
	&R[\mathfrak M_B^{ 0, \bullet}]= 2(N_2-N_1-1)(1-R_{b})+(2N_2-2)(1-R_{\phi_b})+2F(1-R_{Q}) - 2N_2, \\
	&R[\mathfrak M_B^{\bullet, \bullet}]= 2(2N_2-N_1-2)(1-R_{b})+(2N_2-2)(1-R_{\phi_b})+2F(1-R_{Q})- 2(N_2-N_1-1) - 2N_2.
\end{align}
The global symmetry group is $SU(2F)\times U(1)\times U(1)$.
%\be
%\begin{tabular}{c c c c}
% & $SU(2F)$ & $U(1)_{B_1}$ & $U(1)_{B_2}$ \\
%\hline
%$b$ & $\mathbf{1}$ & $-1$ & $0$ \\
%$q$ & $\mathbf{2F}$ & $0$ & $1$ \\
%$\phi_b$ & $\mathbf{1}$ & $2$ & $0$ \\
%$\mathfrak M_B^{\bullet, 0}$ & $\mathbf{1}$ & $2N_2$ & $0$ \\
%$\sigma_B$ & $\mathbf{1}$ & $-2N_2$ & $0$ \\
%$\mathfrak M_B^{0,\bullet}$ & $\mathbf{1}$ & $-2(N_1+N_2-1)$ & $-2F$ \\
%$\mathfrak M_B^{\bullet, \bullet}$ & $\mathbf{1}$ & $-2N_1$ & $-2F$ \\
%\end{tabular}
%\ee
%where, due to the presence of the antisymmetric $\phi_b$ the $U(1)_{B_1}$ charge reads
%\be
%\begin{split}
%	B_1[\mathfrak M^{\vec m, \vec n}]=&-\frac{1}{2}B_1[b]\sum_{a=1}^{N_1}
%	\sum_{b=1}^{N_2}\sum_{\sigma, \gamma=0,1}|(-1)^\sigma m_a + (-1)^\gamma n_b| + \\
%	&-\frac{1}{2}B_1[\phi_b]\sum_{b_1<b_2}\sum_{\sigma, \gamma=0,1}
%	|(-1)^\sigma n_{b_1} + (-1)^\gamma n_{b_2}|
%\end{split}
%\ee
Observe that, even though we have three sets of chiral fields $b, \; \phi_b$ and $Q$ we only have two independent abelian global symmetries due to the superpotential constraint $\Tr(b \phi_b b)$.
%Let us start mapping the generators of the chiral ring between $\mathcal T_A$ and $\mathcal T_B$.
The $R$-charge map with respect to theory $\mathcal T_A$ can be obtained by looking at the mesonic operators. The operator $\Tr(Q Q)$ in $\mathcal T_A$ maps to $\Tr(Q Q)$ in $\mathcal T_B$. 
%being both oh them charged under the flavour symmetry group $SU(2F)$.
%Moreover, we choose the normalisation of the $U(1)_{B_1}$ and $U(1)_{B_2}$ charges of the fundamental fields such that they directly map to the $U(1)_{A_1}$ and $U(1)_{A_2}$ symmetries respectively.
%Before discussing the operator map it is important to compute the $R$-charge of the various monopole operators. This mapping implies the relation $R_Q =R_{q}.\ee
Moreover, the singlet $\Tr(BB)$ in $\mathcal T_A$ maps to $\Tr(\phi_b)$, which gives
\be\label{TBrel2}
	R_{\phi_b}=2 R_B, \qquad R_{b}=1-R_B.
\ee
The chiral ring generators include mesonic operators
\be
	\Tr(Q_{i} Q_{j}), \quad \Tr(\phi_b^J), \quad  \Tr(Q_{i} \phi_b^J Q_{j}),
\ee
for $J=1, \dots, N_2$. The monopoles in the chiral ring include the bare ones $\mathfrak M_B^{0, \bullet}, \, \mathfrak M_B^{\bullet, \bullet}$, while $\mathfrak M_B^{\bullet, 0}$ is removed from the chiral ring by the superpotential term $\sigma_B \,\mathfrak M_B^{\bullet, 0}$, leaving instead $\sigma_B$ as a generator. There are also dressed monopoles as discussed at length in the previous sections. In this particular case we have $\{(\mathfrak M_B^{0, \bullet})_{\phi_b^J}\}$ with 
$J=1, \dots, N_2$. Unlike theory $\mathcal T_A$ there is no $\mathfrak M_B^{\bullet, \bullet}$ dressed with bifundamental chiral $b$ since the superpotential set to zero the matrix $bb$ on the chiral ring.
\subsubsection*{Second dual}
Similarly to what has been done in the previous section, one may apply Aharony duality to the $Sp(N_2)$ node in $\mathcal T_A$ to get the following dual frame  
\be
\scalebox{1}{
\begin{tikzpicture}[baseline]
\tikzstyle{every node}=[font=\footnotesize]
%\node[draw, rectangle] (sqnodeL) at (-4,0) {$2$};
\node[draw=none] (node1) at (0,0) {$Sp(N_1)$};
\draw[black,solid] (node1) edge [out=225,in=135,loop,looseness=5]node[left]{$\phi_c$} (node1) ;
\node[draw=none] (node2) at (4,0) {$Sp(N_1-N_2+F-1)$};
\node[draw, rectangle] (sqnode) at (2,2.5) {$2F$};
\draw[black,solid] (sqnode) edge [out=45,in=135,loop,looseness=5]node[above]{$M$} (sqnode) ;
\draw[black,solid] (node1) edge node[below]{$b$} (node2) ;
\draw[black,solid] (node2) edge node[above]{$q$} (sqnode) ;
\draw[black,solid] (node1) edge node[above]{$p$} (sqnode) ;
\node[draw=none] at (2.2,-1) {$\mathcal W=\sigma_C \, \mathfrak M_C^{0,\bullet} + \Tr(b \phi_c b) + \Tr(bpq)+\Tr(qMq)$};
\node[draw=none] at (-1,1.5) {$\mathcal T_{C}:$};
\end{tikzpicture}}
\ee
In analogy to the case of theory $\mathcal T_B$ there are only two abelian symmetries because of the superpotential terms, giving is $SU(2F) \times U(1)\times U(1)$.
%and the charges are as follows
%\be
%\begin{tabular}{c c c c}
% & $SU(2F)$ & $U(1)_{C_1}$ & $U(1)_{C_2}$ \\
%\hline
%$d$ & $\mathbf{1}$ & $-1$ & $0$ \\
%$p$ & $\mathbf{2F}$ & $1$ & $1$ \\
%$q$ & $\mathbf{2F}$ & $0$ & $-1$ \\
%$M$ & $\mathbf{F(2F-1)}$ & $0$ & $2$ \\
%$\phi_c$ & $\mathbf{1}$ & $2$ & $0$ \\
%$\mathfrak M_C^{\bullet, 0}$ & $\mathbf{1}$ & $-2N_2$ & $0$ \\
%$\mathfrak M_C^{0,\bullet}$ & $\mathbf{1}$ & $2N_1$ & $2F$ \\
%$\sigma_C$ & $\mathbf{1}$ & $-2N_1$ & $-2F$ \\
%$\mathfrak M_C^{\bullet, \bullet}$ & $\mathbf{1}$ & $-2(N_1+N_2-1)$ & $0$ \\
%\end{tabular}
%\ee
The formula for the $R$-charge of a monopole with general flux is completely analogous to \eqref{RchMonB}, so we do not repeat it here. Instead, we can write the $R$-charges of the various minimal monopole operators:
\begin{align}
R[\mathfrak M_C^{\bullet, 0}] &= 2(N_1-N_2+F-1)(1-R_b)+2F(1-R_p)+(2N_1-2)(1-R_{\phi_c}) 	- 2N_1, \\
R[\mathfrak M_C^{ 0, \bullet}] &= 2N_1(1-R_b)+2F(1-R_q)-2(N_1-N_2+F-1), \\
R[\mathfrak M_C^{\bullet, \bullet}]&=2(2N_1-N_2+F-2)(1-R_b)+(2N_1-2)(1-R_{\phi_c})+\notag \\
	&+2F(1-R_q)+2F(1-R_p)-2N_1-2(N_1-N_2+F-1) - 2N_1.
\end{align}

It is possible to get the $R$-charge map with respect to theory $\mathcal T_A$ by looking at mesonic operators. The mesons $\Tr(Q_i Q_j)$ map to $M_{ij}$ so we get
$R_M=2R_Q$ that further gives, from the superpotential term $\Tr(qMq)$, $R_q=1-R_Q$. 
In a similar fashion, the operator $\Tr(BB)$ maps to $\phi_c$ and gives $R_{\phi_c}=2R_B$, that is related, via the superpotential coupling $\Tr(b \phi_c b)$, to $R_b$, implying: $R_b=1-R_B$. The last piece of information comes from the interaction $\Tr(bpq)$, constraining 
the $R$-charge of $p$ to be $R_p=R_B+R_Q$.
The chiral ring generators include the singlets $M_{ij}$ in the antisymmetric representation of the flavour group $SU(2F)$, the powers of the $Sp(N_1)$ antisymmetric $\Tr(\phi_c^J)$ for $J=1,\dots, N_1$ and $\Tr(p_i \phi_c^J p_j)$, for $J=0 ,\dots, N_1$. Furthermore, we have the singlet $\sigma_c$ flipping the monopoles $\mathfrak M_C^{0, \bullet}$. The discussion on dressed monopoles is almost identical to the one in theory $\mathcal T_B$: there are no monopoles 
$\mathfrak M_C^{\bullet, \bullet}$ dressed with the bifundamental $b$ because of $F$-term equations, but there are $\mathfrak M_C^{0, \bullet}$ dressed with powers of $\phi_c^J$: 
$\{(\mathfrak M_C^{\bullet, 0})_{\phi_c^J}\}$, with $J=0, \dots, N_1$.

\subsubsection*{Operator map}
In the previous sections we discussed two dual Aharony-dual frames of the two node quiver theory $\mathcal T_A$. At this point we are ready to study the map of chiral ring operators among the three theories $\mathcal T_A$, $\mathcal T_B$ and $\mathcal T_C$. It is then useful to summarise the $R$-charge map between $\mathcal T_B$, $\mathcal T_A$ and $\mathcal T_C$, $\mathcal T_A$:
\begin{itemize}
\item{$\mathcal T_B \to \mathcal T_A$:
\be	
	R_q=R_Q, \qquad R_b=1-R_B, \qquad R_{\phi_b}=2R_B.
\ee
}
\item{$\mathcal T_C \to \mathcal T_A$:
\be	
	R_q=1-R_Q, \quad R_b=1-R_B, \quad R_p=R_B+R_Q, \quad R_{\phi_c}=2R_B.
\ee
}
Using these maps and the transformation properties of the various operators under the global symmetry it is possible to get the following chiral ring map
\end{itemize}
\be\label{OMSpSp}
\begin{tabular}{c c c c c}
$\mathcal T_A$ & & $\mathcal T_B$ & & $\mathcal T_C$ \\
\hline
$\Tr(Q_i Q_j)$ & & $\Tr(Q_i Q_j)$ & & $M_{ij}$ \\
$\Tr(Q_i (B B)^J Q_j)$ & & $\Tr(Q_i \phi_b^{J} Q_j)$ & & $\Tr(p_i \phi_c^{J-1} p_j)$\\
$\Tr\left((B B)^J\right)$ & & $\Tr(\phi_b^J)$ & & $\Tr(\phi_c^J)$ \\
$\mathfrak M_A^{\bullet, 0}$ & & $\sigma_B$& & $\mathfrak M_C^{\bullet, \bullet}$ \\
$\mathfrak M_A^{0, \bullet}$  & & $\mathfrak M_B^{\bullet, \bullet}$& & $\sigma_C$ \\
$\{(\mathfrak M_A^{\bullet, \bullet})_{(BB)^J}\}$ & & $\{(\mathfrak M_B^{0, \bullet})_{\phi_b^J}\}$& & 
$\{(\mathfrak M_C^{\bullet,0})_{\phi_c^J}\}$ \\
\end{tabular}
\ee
It is interesting to observe how the various (dressed) monopole operators map across the two dual frames: apart from the usual monopole-singlet map coming from the application of Aharony duality, \emph{e.g.} $\mathfrak M_A^{\bullet, 0} \to \sigma_B$ and $\mathfrak M_A^{0, \bullet} \to \sigma_C$, we have monopoles that get ``stretched" across duality, for instance 
$\mathfrak M_A^{0, \bullet} \to \mathfrak M_B^{\bullet, \bullet}$ and 
$\mathfrak M_A^{\bullet, 0} \to \mathfrak M_C^{\bullet, \bullet}$. From the point of view of dressed monopoles this implies that ``long" monopoles dressed with the bifundamental in theory $\mathcal T_A$ map to single-node flux monopoles dressed with antisymmetric fields, as can be seen from the last line of \eqref{OMSpSp}.

Before closing this section, we need to comment the operator map in the following sense. In theory $\mathcal T_A$,
$\Tr\left((BB)^J\right)$ gives algebraically independent operators on the chiral ring for $J=1, \dots, N_1$. The same comment holds for $\Tr(\phi_b^J)$ in theory $\mathcal T_B$, while for $\mathcal T_B$
one has $\Tr(\phi_c^J)$ for $J=1, \dots, N_2$. For this reason, the map involving these operators for $N_1<J \le N_2$ (including monopoles dressed with them) has to be understood such that the operators in $\mathcal T_A$ and $\mathcal T_C$ are composite. Their composite-ness, while evident in theory $\mathcal T_A$, is due to some quantum relation in $\mathcal T_B$ and $\mathcal T_C$.

\subsubsection*{Supersymmetric index}
In this last section we compute the supersymmetric index for the three dual frames $\mathcal T_A$, $\mathcal T_B$ and $\mathcal T_C$ for $N_1=1, \, N_2=3, \, F=3$ and with the choice of $R$-charges given by $R_B=2/5, \, R_Q=2/7$:
\be\label{IndSpSp}
\begin{split}
	\mathcal I&=1+{\color{blue}{28 x^{4/7} y_B^2}}+{\color{blue}{x^{4/5} y_Q^2}}+
	{\color{blue}{\frac{x^{32/35}}{y_B^8 y_Q^2}}}+406 x^{8/7} y_B^4+{\color{blue}{\frac{x^{46/35}}{y_B^8 y_Q^6}}}+{\color{blue}{56 x^{48/35} y_B^2 y_Q^2}}+\frac{28 x^{52/35}}{y_B^6 y_Q^2}+\\
	&+x^{8/5} \left(y_Q^4+{\color{blue}{\frac{1}{y_Q^6}}}\right)+x^{12/7} \left(4060 y_B^6+\frac{1}{y_B^8}\right)+\frac{x^{64/35}}{y_B^{16} y_Q^4}+\frac{28 x^{66/35}}{y_B^6 y_Q^6}+1190 x^{68/35} y_B^4 y_Q^2-65 x^2+\\
	&+\frac{406 x^{72/35}}{y_B^4 y_Q^2}+{\color{blue}{\frac{x^{74/35}}{y_B^8 y_Q^4}}}+x^{76/35} \left(56 y_B^2 y_Q^4+\frac{28 y_B^2}{y_Q^6}\right)+\frac{x^{78/35}}{y_B^{16} y_Q^8}+x^{16/7} \left(31464 y_B^8+\frac{56}{y_B^6}\right)+\dots\\
\end{split}
\ee
where $y_B$ and $y_Q$ are the fugacities for the two abelian symmetries. The charges of the fundamental fields under these symmetries are assigned as follows:
\be
\begin{tabular}{c c c}
$$ & $U(1)_{B}$ & $U(1)_{Q}$ \\
\hline
$B$ & $1$ & $0$ \\ 
$Q$ & $0$ & $1$ \\ 
\hline
$b$ & $-1$ & $0$ \\ 
$Q$ & $0$ & $1$ \\ 
\hline
$b$ & $-1$ & $0$ \\ 
$q$ & $0$ & $-1$ \\ 
\end{tabular}
\ee
In \eqref{IndSpSp} we highlighted in blue the various chiral ring generators already discussed in the previous sections. For definiteness, let us take the generators in the frame $\mathcal T_A$ and identify them with the various terms we obtain in the supersymmetric index
\begin{align}
	\Tr(Q_iQ_j) \;&\leftrightarrow\; {\color{blue}{28 x^{4/7} y_B^2}} \\
	\Tr(BB) \;&\leftrightarrow \;{\color{blue}{x^{4/5} y_Q^2}} \\
	\Tr(Q_i(BB)Q_j) \;&\leftrightarrow \;{\color{blue}{56 x^{48/35} y_B^2 y_Q^2}} \\
	\mathfrak M_A^{\cdot, 0} \;&\leftrightarrow \; {\color{blue}{\frac{x^{8/5}}{y_Q^6}}} \\
	\mathfrak M_A^{0, \cdot} \;&\leftrightarrow \;  {\color{blue}{\frac{x^{32/35}}{y_B^8 y_Q^2}}} \\
	\mathfrak M_A^{\cdot, \cdot} \;&\leftrightarrow \; {\color{blue}{\frac{x^{46/35}}{y_B^8 y_Q^6}}} 
	\\
	\{(\mathfrak M_A^{\cdot, \cdot})_{BB} \}\;&\leftrightarrow
	{\color{blue}{\frac{x^{74/35}}{y_B^8 y_Q^4}}}.
\end{align}

%%%%%%%%%%%%%%%%%%%%%%%%%%%%%%%%%%%%%%%%%%%%%%%%%%%%%%%%%%%%%%%
%%%%%%%%%%%%%%%%%%%%%%%%%%%%%%%%%%%%%%%%%%%%%%%%%%%%%%%%%%%%%%%
%%%%%%%%%%%%%%%%%%%%%%%%%%%%%%%%%%%%%%%%%%%%%%%%%%%%%%%%%%%%%%%
%%%%%%%%%%%%%%%%%%%%%%%%%%%%%%%%%%%%%%%%%%%%%%%%%%%%%%%%%%%%%%%
%%%%%%%%%%%%%%%%%%%%%%%%%%%%%%%%%%%%%%%%%%%%%%%%%%%%%%%%%%%%%%%
\section{Dualities in quivers with baryons and baryon-monopoles}\label{typeII}
In this section we study quivers with more general gauge groups, including $SO$ and $SU$ gauge groups. We work out three examples, with gauge groups $SO (N_1) \times Sp(N_2)$, $SO (N_1) \times SO(N_2)$ and $Sp (N_1) \times SU(N_2)$. In each example we are able to find the mapping of all the chiral ring generators. 

Unlike the cases discussed in section \ref{typeI}, for the examples in this section we do not have a general rule for the mapping of the monopoles which can be extended to quivers with an arbitrary number of gauge groups. We leave this issue to future work.

\subsection{$SO (N_1) \times Sp(N_2)$ theory}\label{SOSP}
Let us consider the following theory:
\be
\label{pic:TASOSp}
\scalebox{1}{
\begin{tikzpicture}[baseline]
\tikzstyle{every node}=[font=\footnotesize]
%\node[draw, rectangle] (sqnodeL) at (-4,0) {$2$};
\node[draw=none] (node1) at (0,0) {$SO(N_1)$};
\node[draw=none] (node2) at (3,0) {$Sp(N_2)$};
\node[draw, rectangle] (sqnode) at (6,0) {$F$};
\draw[black,solid] (node1) edge node[above]{$B$} (node2) ;
\draw[black,solid] (node2) edge node[above]{$Q$} (sqnode) ;
%\draw[black,solid,->] (node2) edge [out=25,in=155,loop,looseness=1]node[above]{$F$} (sqnodeR) ;
\node[draw=none] at (3,-1) {$\mathcal W=0$};
\node[draw=none] at (-2,0) {$\mathcal T_{A}:$};
%node[above]{\color{blue}$X$} node[below, yshift=0cm] {$\red F_X$} node {$ \red \times$}  (node);
\end{tikzpicture}}
\ee
with $N_1+F$ even. In order to avoid runaway superpotential breaking supersymmetry to be generated or confining dynamics, we assume $2N_2\geq N_1$ and $N_1+F> 2N_2+2$. 

The continuous global symmetries of the model are $SU(F)\times U(1)_A\times U(1)_F$, where $U(1)_A$ acts on the bi-fundamental fields and $U(1)_F$ acts on the flavor fields. Moreover, two discrete factors, the charge-conjugation $\mathbb{Z}_2^\cC$ and the magnetic symmetry $\mathbb{Z}_2^\cM$, are associated to the orthogonal node. The charge-conjugation acts on flavor and bi-fundamental chirals as an orthogonal reflection transformation and also possesses a non-trivial action on monopoles; in particular, as reviewed in section \ref{sec:reviewSO}, two kind of monopoles can be defined, whose fluxes are denoted by $\pm$, {\it i.e.} their charge under $\cC$. The magnetic symmetry $\cM$ instead, is related to center of the orthogonal group and acts trivially on bi-fundamental and chiral fields while it charges $-1$ all the monopoles with minimal flux.   

Let us study the chiral ring generators of this model. Because of the presence of an $SO(N_1)$ gauge group, we have both mesonic and baryonic operators. The first kind consists of traces involving both the bi-fundamental field $B$ and the flavors $Q_i$:
\be
\begin{split}
&{\Tr}\left((BB)^{2j}\right)\,,j=1,\dots,\lfloor N_1/2 \rfloor \\
&{\Tr}\left(Q_i\,(BB)^{J}\,Q_j\right)\,, J=0,\dots,2N_2-1\,.
\end{split}
\ee
Observe that, because the orthogonal invariant form is symmetric and the symplectic one is antisymmetric, it is not possible to construct a meson operator that is quadratic in the bi-fundamental fields.
The unique baryon operator is:
\be
\varepsilon B^{N_1}{\color{red}Q_i}\,=\,\varepsilon_{a_1\,\dots\,a_{2r}\color{red}{a_{2r+1}}}\,(B^{a_1}\cdot B^{a_2})\cdots (B^{a_{2r-1}}\cdot B^{a_{2r}})\,{\color{red}(B^{a_{2r+1}}\cdot Q_i)}\,,
\ee
where $\cdot$ denotes the contraction of the $Sp(N_2)$ indices through the symplectic form and $r=\lfloor N_1/2\rfloor$ is the rank of the orthogonal group. The term colored in red is only present when $N_1$ is odd, in which case we need an extra flavor in order to obtain a gauge invariant operator: this implies that when $N_1$ is odd, such baryon transforms non-trivially under $SU(F)$.\footnote{Other baryonic-like operators can be obtained in principle contracting pairs of bi-fundamentals with flavors rather that with the symplectic form; however, these are composite operators.} As reviewed in appendix \ref{sec:reviewSO}, also the monopoles operators can be of non-baryonic and of baryonic type. The non-baryonic monopoles are charge-conjugation even and consists of $\mathfrak{M}_A^{+,0},\,\mathfrak{M}_A^{0,\bullet}$ and $\mathfrak{M}_A^{+,\bullet}$.  $\mathfrak{M}_A^{+,\bullet}$, with minimal fluxes turned for both the nodes, can be dressed with powers of bi-fundamental fields, $(\mathfrak{M}_A^{+,\bullet})_{(BB)^j}$, $j=0,\dots,\lfloor N_1/2\rfloor$. The baryonic monopoles, instead, are charge-conjugation odd and always have non-trivial fluxes turned on for the orthogonal group. They need to be dressed with $N_1-2$ chiral fields in the fundamental representation of $SO(N_1)$; in the case at hand we have two possibilities:
\be
(\mathfrak{M}_A^{-,0})_{B^{N_1-2}{\color{red}Q_i}}\,,\quad ( \mathfrak{M}_A^{-, \bullet})_{B^{N_1-2}{\color{red}Q_i}}\,.
\ee
As before, the $Sp(N_2)$ indices of the dressing factors are contracted with the symplectic form and possibly with an extra chiral field $Q_i$ if $N_1$ is odd; in this latter case, both the baryon monopoles transform in the fundamental representation of $SU(F)$. The R-charges of the various monopoles can be computed using the general formula:
\be \label{monTA-SOSp}
\begin{split}
	&R[\mathfrak M_A^{\vec m, \vec n}]=\\
	&\frac{1}{2} (1-R_B) \sum_{a=1}^{\lfloor N_1/2\rfloor}\sum_{b=1}^{N_2}
	\sum_{\sigma, \gamma={0,1}} |(-1)^\sigma m_a+(-1)^\gamma n_b|+{\color{red}\frac{1}{2} (1-R_B) \sum_{b=1}^{N_2}
	\sum_{\gamma={0,1}} |(-1)^\gamma n_b|}\\
	&+\frac{1}{2} F(1-R_Q)\sum_{b=1}^{N_2}\sum_{\gamma=0,1} |(-1)^\gamma n_b|-\sum_{b_1<b_2}^{N_2}\sum_{\gamma=0,1}|n_{b_1}+(-1)^\gamma n_{b_2}|
	-\sum_{b=1}^{N_2} |2n_b|+\\
	&-\sum_{a_1<a_2}^{\lfloor  N_1/2\rfloor}\sum_{\sigma=,0,1} |m_{a_1}+(-1)^\sigma m_{a_2}|\,{\color{red}-\sum_{a}^{\lfloor  N_1/2\rfloor}\sum_{\sigma=,0,1} |(-1)^\sigma m_{a}|}\,,
\end{split}
\ee
where the red terms must be taken into account whenever $N_1$ is odd. Specializing the previous formula to the case of the monopole with minimal fluxes, we obtain the following result:
\be
\begin{split}
&R[\mathfrak{M}^{+,0}_A]\,=\, 2N_2(1-R_B)\,-\,(N_1-2)\,,\\
&R[\mathfrak{M}^{0,\bullet}_A]\,=\,F(1-R_Q)\,+\,N_1(1-R_B)\,-\,2N_2\,,\\
&R[\mathfrak{M}^{+,\bullet}_A]\,=\,F(1-R_Q)\,+\,(2N_2+N_1-2)(1-R_B)\,-\,(2N_2+N_1-2)\,.
\end{split}
\ee
 Since $R[\mathfrak{M}^{+,0}_A]=R[\mathfrak{M}^{-,0}_A]$ and $R[\mathfrak{M}^{+,\bullet}_A]=R[\mathfrak{M}^{-,\bullet}_A]$, it is straightforward to obtain the baryon-monopole R-charges taking into account the contribution of the dressings.
 
\subsubsection*{First dual}
Using the ASRW duality reviewed in appendix \ref{sec:reviewSO}, one can dualize the orthogonal node in \eqref{pic:TASOSp}, obtaining the following dual frame:
\be
\scalebox{1}{
\begin{tikzpicture}[baseline]
\tikzstyle{every node}=[font=\footnotesize]
%\node[draw, rectangle] (sqnodeL) at (-4,0) {$2$};
\node[draw=none] (node1) at (0,0) {$SO(2N_2-N_1+2)$};
\node[draw=none] (node2) at (4,0) {$Sp(N_2)$};
\node[draw, rectangle] (sqnode) at (7,0) {$F$};
\draw[black,solid] (node1) edge node[above]{$b$} (node2) ;
\draw[black,solid] (node2) edge node[above]{$Q$} (sqnode) ;
\draw[black,solid] (node2) edge [out=45,in=135,loop,looseness=5]node[above]{$S$} (node2) ;
%\draw[black,solid,->] (node2) edge [out=25,in=155,loop,looseness=1]node[above]{$F$} (sqnodeR) ;
\node[draw=none] at (4,-1) {$\mathcal W=\sigma_B \, \mathfrak M_B^{+, 0} + \text{Tr}(bSb)$};
\node[draw=none] at (-3,0) {$\mathcal T_{B}:$};
\end{tikzpicture}}\qquad \qquad
\ee 
where $S$ is a symmetric (adjoint) field. As before, the chiral ring contains both non-baryonic and baryonic operators. Among the possible traces, all the ones containing the symmetric product $bb$ are flipped while the traces involving the adjoint field
 \be
 {\Tr}(Q_iS^{J}Q_j)\,,\,J=1,\dots 2N_2-1\,, \qquad {\Tr}S^{2j}\,,\,j=1, \dots N_2\,,
 \ee
 are now part of the chiral ring. The unique baryon is instead $\varepsilon(b^{\tilde N_1} {\color{red}Q_i})$ where $\tilde N_1=2N_2-N_1+2$. As usual, the operators ${\Tr}(BB)^2$ and ${\Tr}Q_iQ_j$ in $\cT_A$ map to ${{\Tr}\,S^2}$  and ${\Tr}Q_iQ_j$ respectively in $\cT_B$. From this map we immediately read the constraints on the R-charges of $b$, $S$ and $Q_i$ in order for the duality to hold:
 \be
R_S/2\,=\,1-R_b\,=\, R_B\,.
 \ee
 The spectrum of the conjugation-even monopole generators are $\mathfrak{M}_B^{+,0}, \mathfrak{M}_B^{+,\bullet}$ and $\left\{(\mathfrak{M}_B^{0,\bullet})_{S^J}\right\}$. The baryonic monopoles are instead $(\mathfrak{M}_B^{-,0})_{b^{\tilde N_1}{\color{red} Q_i}}$ and $(\mathfrak{M}_B^{-,\bullet})_{b^{\tilde N_1}{\color{red} Q_i}}$, following the same way of reasoning explained in the case of $\mathcal{T}_A$. The R-charge of such operators can be computed using the general formula:
\be \label{monTB-SOSp}
\begin{split}
	&R[\mathfrak M_A^{\vec m, \vec n}]=\\
	&\frac{1}{2} (1-R_b) \sum_{a=1}^{\lfloor N_1/2\rfloor}\sum_{b=1}^{N_2}
	\sum_{\sigma, \gamma={0,1}} |(-1)^\sigma m_a+(-1)^\gamma n_b|+{\color{red}\frac{1}{2} (1-R_b) \sum_{b=1}^{N_2}
	\sum_{\gamma={0,1}} |(-1)^\gamma n_b|}\\
	&+\frac{1}{2} F(1-R_Q)\sum_{b=1}^{N_2}\sum_{\gamma=0,1} |(-1)^\gamma n_b|-\sum_{b_1<b_2}^{N_2}\sum_{\gamma=0,1}|n_{b_1}+(-1)^\gamma n_{b_2}|
	-\sum_{b=1}^{N_2} |2n_b|+\\
	&+(1-R_S)\sum_{b_1<b_2}^{N_2}\sum_{\gamma=0,1}|n_{b_1}+(-1)^\gamma n_{b_2}|
	+(1-R_S)\sum_{b=1}^{N_2} |2n_b|+\\
	&-\sum_{a_1<a_2}^{\lfloor N_1/2\rfloor}\sum_{\sigma=,0,1} |m_{a_1}+(-1)^\sigma m_{a_2}|\,{\color{red}-\sum_{a}^{\lfloor N_1/2\rfloor}\sum_{\sigma=,0,1} |(-1)^\sigma m_{a}|}\,,
\end{split}
\ee
In particular, the R-charges of the charge-conjugation even monopoles are:
 \be
 \label{eq:RchargesTBsoUsp}
 \begin{split}
 &R[\mathfrak{M}^{+,0}_B]\,=\,2N_2\,(1-R_b)-(2N_2-N_1)\,,\\
  &R[\mathfrak{M}^{0,\bullet}_B]\,=\,F\,(1-R_Q)\,+\,(2N_2-N_1+2)(1-R_b)\,+\,2N_2\,(1-R_S)-2N_2\,,\\
  &R[\mathfrak{M}^{+,\bullet}_B]\,=\,F\,(1-R_Q)\,+\,(4N_2-N_1)(1-R_b)+2N_2(1-R_S)-(4N_2-N_1)\,.
 \end{split}
 \ee

\subsubsection*{Second dual}
This time we apply Aharony duality to the symplectic node in $\cT_A$:
\be
\scalebox{1}{
\begin{tikzpicture}[baseline]
\tikzstyle{every node}=[font=\footnotesize]
%\node[draw, rectangle] (sqnodeL) at (-4,0) {$2$};
\node[draw=none] (node1) at (0,0) {$SO(N_1)$};
\draw[black,solid] (node1) edge [out=225,in=135,loop,looseness=5]node[left]{$A$} (node1) ;
\node[draw=none] (node2) at (4,0) {$Sp\left(\frac{F+N_1}{2}-N_2-1\right)$};
\node[draw, rectangle] (sqnode) at (2,2.5) {$F$};
\draw[black,solid] (sqnode) edge [out=45,in=135,loop,looseness=5]node[above]{$M$} (sqnode) ;
\draw[black,solid] (node1) edge node[below]{$\tilde b$} (node2) ;
\draw[black,solid] (node2) edge node[above]{$\tilde q$} (sqnode) ;
\draw[black,solid] (node1) edge node[above]{$p$} (sqnode) ;
\node[draw=none] at (2.2,-1) {$\mathcal W=\sigma_C \, \mathfrak M_C^{0,\bullet} + \Tr(\tilde b A \tilde b) + \Tr(\tilde bp \tilde q)+\Tr(\tilde q M \tilde q)$};
\node[draw=none] at (-1,1.5) {$\mathcal T_{C}:$};
\end{tikzpicture}}
\ee
where $A$ transforms in the rank-two antisymmetric (adjoint) representation of $SO(N_1)$ and $M$ transforms in the rank-two antisymmetric representation of the $SU(F)$ flavor group. Because $\text{Tr}(BB)^2$ and $Q_i Q_j$ in $\cT_A$ map to $\text{Tr}\,A^2$ and $M_{ij}$ respectively in $\cT_C$, the $R$-charges of the chiral fields in this third frame read:
\be
R_A\,=\,2(1-R_{\tilde{b}})\,=\,2R_B\,,\quad R_M\,=\,2(1-R_{\tilde{q}})\,=\,2R_Q\,,\quad R_p\,=\,R_B+R_Q\,.
\ee
Because of the constraints imposed by the superpotential, a unique baryon can be built using $A$ (and $p$ whenever $N_1$ is odd) as follows:
\be
\label{eq:BaryonTCSOSp}
\varepsilon A^{n}{\color{red}p_i}\,=\,\varepsilon_{a_1\,\dots\,a_{2r}\,\color{red}{a_{2r+1}}}\,(A^{a_1\,a_2})\cdots (A^{a_{2r-1}\,a_{2r}})\,\,{\color{red} p^{a_{2r+1}}_i}\,,
\ee
where we remind $r=\lfloor N_1/2\rfloor$ to be the rank of the orthogonal group. As in $\cT_A$, one can wonder whether more general baryons, built substituting one or more adjoint fields with some antisymmetric combination $p_i p_j$ in \eqref{eq:BaryonTCSOSp}, are chiral ring generators. As we will see, the mapping between $\cT_A$ and $\cT_C$ suggests that such baryonic-like objects are actually composite operators that can be expressed as products of the baryon \eqref{eq:BaryonTCSOSp} and the meson $M_{ij}$ due to quantum relations. Similarly, the baryonic-monopoles generators in such frame are $(\mathfrak{M}_{C}^{-,0})_{A^{n-1}{\color{red}p_i}}$ and $(\mathfrak{M}_{C}^{-,\bullet})_{A^{n-1}{\color{red}p_i}}$
The $R$-charges of the fundamental charge-conjugation even monopoles are:
\be
\begin{split}
&R[\mathfrak{M}^{+,0}_C]\,=\,(F+N_1-2N_2-2)(1-R_{\tilde b})\,+\,(N_1-2)(1-R_A)\,+\,F(1-R_p)-(N_1-2)\,,\\
&R[\mathfrak{M}^{0,\bullet}_C]\,=\,N_1(1-R_{\tilde b})\,+\,F(1-R_{\tilde q})-(F+N_1-2N_2-2)\,,\\
&R[\mathfrak{M}^{+,\bullet}_C]\,=\,-R_{\tilde b}\,(2N_1+F-2N_2-4)\,+\,F(2-R_p-R_{\tilde q})\,+\,(N_1-2)(1-R_A)\,.
\end{split}
\ee

\subsubsection*{Operator map}
In this section we want to present the mapping between the chiral ring generators in the three dual frames. The proposal is supported by the agreement between the global charges of the various operators and further checked using the supersymmetric index. The map is:
 \be\label{OMSOSp}
\begin{tabular}{c c c c c}
$\mathcal T_A$ & & $\mathcal T_B$  && $\mathcal T_C$\\
\hline
$\mathfrak M_A^{+, 0}$ & & $\sigma_B$        && $\mathfrak M_C^{+, \bullet}$\\
$\mathfrak M_A^{0, \bullet}$  & & $\mathfrak M_B^{+, \bullet}$        & &    $\sigma_C$\\
$(\mathfrak M_A^{+, \bullet})_{(BB)^J}$  & & $(\mathfrak M_B^{0, \bullet})_{S^J}$         & &    $(\mathfrak M_C^{+, 0})_{A^{J}}$\\
$(\mathfrak M^{-,0}_A)_{B^{N_1-2}{\color{red}Q_i}}$ & & $\varepsilon b^{N_1}{\color{red}Q_i}$      & &    $(\mathfrak{M}_{C}^{-,\bullet})_{A^{r-1}{\color{red}p_i}}$\\
$(\mathfrak M^{-,\bullet}_A)_{B^{N_1-2}{\color{red}Q_i}}$ & & $(\mathfrak M^{-,\bullet}_B)_{b^{N_1-2}{\color{red}Q_i}}$     & &    $(\mathfrak{M}_{C}^{-,0})_{A^{r-1}{\color{red}p_i}}$\\
$\varepsilon B^{N_1}{\color{red}Q_i}$ & & $(\mathfrak M^{-,0}_B)_{b^{N_1-2}{\color{red}Q_i}}$     & &    $\varepsilon A^{r}{\color{red}p_i}$\\
$Q_i Q_j$ & & $Q_i Q_j$      & &    $M_{ij}$\\
$\Tr\left((B B)^{2j}\right)$ & & $\Tr\left(S^{2j} \right)$     & &    $\Tr\left(A^{2j} \right)$\\
$\Tr(Q_i (B B)^{J} Q_j)$ & & $\Tr(Q_i\,S^{J}\,Q_j)$    &&       $\Tr(p_i\,A^{J-1}\,p_j)$
\end{tabular}
\ee
As in the ARSW duality, the baryon monopole $(-,0)$ is mapped to the baryon (and vice-versa) when the orthogonal group is dualized. Also observe that the charge-conjugation even monopoles still behave in agreement with the prescription presented in the introduction. The charge conjugation-odd monopoles behave in a way similar to the the general rule under $Sp$-duality: $\mathfrak{M}^{-,0}$ extends while $\mathfrak{M}^{-,\bullet}$ shortens.

\subsubsection*{Supersymmetric index}
Let us compute a the supersymmetric index of the three dual theories in the particular case of $N_1=3$, $N_2=2$ and $F=7$. We fix the $R$-charges to be $R_B=\frac{2}{5}$ and $R_Q=\frac{1}{2}$; moreover, the charges under the $U(1)$ global symmetries are:
\be
\label{eq:chU(1)SpSO}
\begin{tabular}{c c c}
$$ & $U(1)_{B}$ & $U(1)_{Q}$ \\
\hline
$B$ & $1$ & $0$ \\ 
$Q$ & $0$ & $1$ \\ 
\hline
$b$ & $-1$ & $0$ \\ 
$Q$ & $0$ & $1$ \\ 
\hline
$\tilde b$ & $-1$ & $0$ \\ 
$\tilde q$ & $0$ & $-1$ \\ 
\end{tabular}
\ee
while the charges of mesons and rank-two field are constrained by the form of the superpotential. We denote with $y_B$ and $y_Q$ the fugacities of $U(1)_B$ and $U(1)_Q$ respectively, while we denote with $x$ the $R$-charge fugacity and with $z_i$ the $SU(F)$ ones. With the mentioned choices, the supersymmetric index reads:
\be
\begin{split}
\mathcal{I}=&1\,+\,{\color{blue} 21\,x\,y^2_{Q}}\,+\,{\color{blue} \frac{x^{13/10}}{y_Q^7\,y_B^3}}\,+\,{\color{blue} \frac{x^{7/5}}{y_B^4}}\,+\,{\color{blue} \frac{x^{3/2}}{y_Q^7\,y_B^5} }\,+\,{\color{blue} x^{8/5}\,y_B^4}\,+\,{\color{blue} 7\,x^{17/10}\,y_Q\,y_B^3}\,+\,{\color{blue} 28\,x^{9/5}\,y_B^2\,y_Q^2}\,+\,\\
&+x^2(231\,y_Q^4-50)+x^{23/10}\left( {\color{blue}\frac{1}{y_Q^7\,y_B^3}}+\frac{21}{y_Q^5\,y_B^3}+{\color{blue}\frac{7\,y_Q}{y_B^3}} \right)+x^{12/5}\left({\color{blue}\frac{7}{y_Q^6\,y_B^4}}+\frac{21\,y_Q^2}{y_B^4} \right)+\dots
\end{split}
\ee
We highlighted in blue the generators of the chiral ring. Let us explicitly write down the generators we read from the index:
\begin{align}
\nonumber
	\Tr(Q_iQ_j) \;&\leftrightarrow\; {\color{blue}21\,x\,y_Q^2} & 	\text{Tr}(BB)^2 \;&\leftrightarrow\; {\color{blue}x^{8/5}\,y_B^4}   \\
\nonumber
	\text{Tr}(Q_i(BB)Q_j) \;&\leftrightarrow\; {\color{blue}28\,x^{9/5}\,y_Q^2\,y_B^2}  & 	\varepsilon B^{N_1}{Q_i} \;&\leftrightarrow\; {\color{blue}7\,x^{17/10}\,y_Q\,y_B^3}\\
\nonumber
	\mathfrak{M}_A^{0,\bullet} \;&\leftrightarrow\; {\color{blue}\frac{x^{13/10}}{y_Q^7\,y_B^3}} & \mathfrak{M}_A^{+,0} \;&\leftrightarrow\; {\color{blue}\frac{x^{7/5}}{y_B^4}} \\
\nonumber
	\mathfrak{M}_A^{+,\bullet} \;&\leftrightarrow\; {\color{blue}\frac{x^{3/2}}{y_Q^7\,y_B^5}}  & 	(\mathfrak{M}_A^{+,\bullet})_{(BB)} \;&\leftrightarrow\; {\color{blue}\frac{x^{23/10}}{y_Q^7\,y_B^3}} \\
\nonumber	
	(\mathfrak M^{-,0}_A)_{B^{N_1-2}Q_i} \;&\leftrightarrow\; {\color{blue}\frac{7\,x^{23/10}\,y_Q}{y_B^3}}  & (\mathfrak M^{-,\bullet}_A)_{B^{N_1-2}{Q_i}} \;&\leftrightarrow\; {\color{blue}\frac{7\,x^{12/5}}{y_Q^6\,y_B^4}} \\
\end{align}
%%%%%%%%%%%%%%%%%%%%%%%%%%%%%%%%%%%%%%%%%
%%%%%%%%%%%%%%%%%%%%%%%%%%%%%%%%%%%%%%%%%
%%%%%%%%%%%%%%%%%%%%%%%%%%%%%%%%%%%%%%%%%
%%%%%%%%%%%%%%%%%%%%%%%%%%%%%%%%%%%%%%%%%

\subsection{$SO (N_1) \times SO(N_2)$ theory}\label{SOSO}

Let us consider a theory involving orthogonal gauge groups only:
\be
\scalebox{1}{
\begin{tikzpicture}[baseline]
\tikzstyle{every node}=[font=\footnotesize]
%\node[draw, rectangle] (sqnodeL) at (-4,0) {$2$};
\node[draw=none] (node1) at (0,0) {$SO(N_1)$};
\node[draw=none] (node2) at (3,0) {$SO(N_2)$};
\node[draw, rectangle] (sqnode) at (6,0) {$F$};
\draw[black,solid] (node1) edge node[above]{$B$} (node2) ;
\draw[black,solid] (node2) edge node[above]{$Q$} (sqnode) ;
%\draw[black,solid,->] (node2) edge [out=25,in=155,loop,looseness=1]node[above]{$F$} (sqnodeR) ;
\node[draw=none] at (3,-1) {$\mathcal W=0$};
\node[draw=none] at (-2,0) {$\mathcal T_{A}:$};
%node[above]{\color{blue}$X$} node[below, yshift=0cm] {$\red F_X$} node {$ \red \times$}  (node);
\end{tikzpicture}}
\ee
We assume $N_2\geq N_1$ and $N_1+F\geq N_2$ in order to avoid the generation of a runaway superpotential or confining dynamics. In the following it will be useful to define the rank of the two groups, $r_i=\lfloor N_i \rfloor$. The continuous global symmetry group is $SU(F)\times U(1)_B\times U(1)_Q$, where the labels of the two Abelian factors denote the field which they act on. The theory also enjoys charge-conjugation symmetry $\mathbb{Z}_2^{\cC_i}$ and magnetic symmetry $\mathbb{Z}_2^{\cM_i}$ for each node. Let us understand the possible generators of the chiral ring. We first mention mesons and baryons. The mesonic operators are as usual:
 \be
 \begin{split}
 \text{Tr}(BB)^J\,,\quad &J=1, \dots, N_1\\
 \text{Tr}(Q_i (BB)^J Q_J)\,, \quad &J=0,\dots, N_1-1\,. 
 \end{split}
 \ee
It is also possible to build three kind of baryons:
\be
\begin{split}
\label{eq:BayonsSOSO}
\varepsilon_1\varepsilon_2B^{N_1}Q^{N_2-N_1}\,&=\,\varepsilon_{a_1\,\dots a_{N_1}} \varepsilon^{b_1\,\dots\,b_{N_2}}\,B^{a_1}_{b_1}\cdots B^{a_{N_1}}_{b_{N_1}}\,Q^{i_1}_{b_{N_1+1}}\cdots Q^{i_{N_2-N_1}}_{b_{N_2}}\,,\\
\varepsilon_1B^{N_1}Q^{N_1}\,&=\,\varepsilon_{a_1\,\dots a_{N_1}} \,B^{a_1}_{b_1}\cdots B^{a_{N_1}}_{b_{N_1}}\,Q_{i_1}^{b_{1}}\cdots Q_{i_{N_1}}^{b_{N_1}}\,,\\
\varepsilon_2{Q}^{N_2}\,&=\,\varepsilon_{b_1\,\dots b_{N_2}}\,Q_{i_1}^{b_{1}}\cdots Q_{i_{N_2}}^{b_{N_2}}\,,
\end{split}
\ee
where $a_k$, $b_k$ and $i_k$ are $SO(N_1)$, $SO(N_2)$ and $SU(F)$ indices respectively. The three baryon operators have charges $(-,-)$, $(-,+)$ and $(+,-)$ respectively under  $(\cC_1, \cC_2)$; they are invariant, instead, with respect to both the magnetic symmetries. Observe that the $(-,-)$ baryon can be always built because by assumption $F\geq N_2-N_1$. The $(-,+)$ operator only exists if $F\geq N_1$ while the last composite operator in \eqref{eq:BayonsSOSO} can be constructed whenever $F\geq N_2$.

The set of the monopole chiral-ring generators is more intricate in this case and it can be arranged in terms of the charges under the $\cC_i$ symmetries. To be more precise, we will consider the bare charge under charge-conjugation: for instance, a monopole such as $\mathfrak{M}^{0,+}$, which at the bare level is even with respect to both charge conjugation symmetries, at the quantum level can transform non-trivially under $\cC_1$, because of the dressing of bi-fundamental fields. The even monopoles (at the bare level as just explained) are $\mathfrak M^{+,0},\, \mathfrak M^{0,+} $ and $\mathfrak M^{+,+}$, where the latter can be dressed with powers of $BB$. Using the standard formulas introduced in the previous sections, the $R$-charges of the monopoles can be easily computed:
\be
\begin{split}
&R[\mathfrak{M}^{+,0}]\,=\,N_2(1-R_B)-(N_1-2)\,,\\
&R[\mathfrak{M}^{0,+}]\,=\,F(1-R_Q)+N_1(1-R_B)-(N_2-2)\,,\\
&R[\mathfrak{M}^{+,+}]\,=\,F(1-R_Q)+(N_2+N_1-2)(1-R_B)-(N_1-2)-(N_2-2)\,.
\end{split}
\ee
As already anticipated, the set of baryonic monopole operators is much more involved. The theory posses three different kinds of baryon monopoles, with bare conjugation charges $(-,+)$, $(+,-)$ or $(-,-)$:
\begin{itemize}
\item Monopoles with charges $(-,+)$ under $(\cC_1, \cC_2)$ can be obtained appropriately dressing $\mathfrak{M}^{-,0}$ or $\mathfrak{M}^{-,+}$. The former can be dressed in two different ways. Schematically: 
\be
\label{eq:barymonSOSO1}
\begin{split}
(\mathfrak{M}^{-,0}_A)_{B^{N_1-2}\,Q^{N_1-2}}\,&\approx\, \mathfrak{M}^{-,0}_A \cdot \varepsilon^{(1)}_{a_1\dots a_{N_1-2}}(B^{a_1}_{b_1}Q^{b_1}_{i_1})\cdots(B^{a_{N_1-2}}_{b_{N_1-2}}Q^{b_{N_1-2}}_{i_{N_1-2}})\\ 
(\varepsilon_2\mathfrak{M}^{-,0}_A)_{B^{N_1-2}Q^{N_2-N_1+2}}\,&\approx \mathfrak{M}^{-,0}_A\cdot\varepsilon^{(1)}_{a_1\dots a_{N_1-2}}\varepsilon_{(2)}^{b_1\dots b_{N_2}}B^{a_1}_{b_1}\cdots B^{a_{N_1-2}}_{b_{N_1-2}}\,Q_{b_{N_1-1}}^{i_{N_1-1}}\cdots Q_{b_{N_1}}^{i_{N_2}} \,\,.
\end{split}
\ee
The two dressings are possible only if $F\geq N_1-2$ and $F\geq N_2-N_1+2$ respectively, because full antisymmetrizations of indices are needed. In the dressing of the second operator in \eqref{eq:barymonSOSO1} the Levi-Civita tensor of the full $SO(N_2)$ node appears, making this monopole someway exotic: it behaves like a baryon with respect to the second node. For this reason, we dub this operator {\it dibaryon monopole}.   In the same way, we can construct:
\be
(\mathfrak{M}^{-,+}_A)_{B^{N_1-2}\,Q^{N_1-2}}\,\approx\, \mathfrak{M}^{-,+}_A \cdot \varepsilon^{(1)}_{a_1\dots a_{N_1-2}}(B^{a_1}_{b_1}Q^{b_1}_{i_1})\cdots(B^{a_{N_1-2}}_{b_{N_1-2}}Q^{b_{N_1-2}}_{i_{N_1-2}})\,.
\ee
Observe that $\mathfrak{M}^{-,+}$ cannot be dressed using the Levi-Civita tensor of the second node. Indeed, when it gets a VEV, the non-trivial $+$ flux causes the breaking of the $SO(N_2)$ gauge group down to $S(O(N_2-2)\times O(2))$ and the presence of the $\varepsilon_{(2)}$ symbol would make the operator not invariant with respect to the residual gauge group.

\item In a similar fashion, we can consider monopoles which are, at the bare level, charge-conjugation odd with respect to $\cC_2$ only. These can be built appropriately dressing $\mathfrak{M}_A^{0,-}$ and $\mathfrak{M}_A^{+,-}$. In the former case we can build:
\be
\begin{split}
(\mathfrak{M}^{0,-}_A)_{Q^{N_2-2}}\,&\approx\, \mathfrak{M}_A^{-,0}\cdot \varepsilon_{(2)}^{b_1\dots b_{N_2-2}}\,Q_{b_1}^{i_1}\cdots Q_{b_{N_2-2}}^{i_{N_2-2}}\,,\\ 
({\varepsilon_1\mathfrak{M}}^{0,-}_A)_{\,B^{N_1}\,Q^{N_2-N_1-2}}\,&\approx\, \mathfrak{M}_A^{-,0}\cdot \varepsilon^{(1)}_{a_1\,\dots a_{N_1}}B^{a_1}_{b_1}\cdots B^{a_{N_1}}_{b_{N_1}}\varepsilon_{(2)}^{b_1\dots b_{N_2-2}}\,Q^{b^{N_1+1}}_{b_{N_1+1}}\cdots Q^{b^{N_2-2}}_{b_{N_2-2}}\,,
\end{split}
\ee
The two baryon monopoles only exists if $F\geq N_2-2$ and $N_1\leq N_2-2$ respectively. The latter, in particular, is a dibaryon monopole in the notation introduced previously. In a similar manner, starting from $\mathfrak{M}_A^{+,-}$ we can build
\be
(\mathfrak{M}^{+,-}_A)_{Q^{N_2-2}}\,\approx\, \mathfrak{M}_A^{-,+}\cdot \varepsilon_{(2)}^{b_1\dots b_{N_2-2}}\,Q_{b_1}^{i_1}\cdots Q_{b_{N_2-2}}^{i_{N_2-2}}\,,
\ee
As before, we cannot use $\varepsilon_{(1)}$ in the dressing of $\mathfrak{M}^{(+,-)}$ because it would make the dressed monopole transforming under the residual gauge group.

\item Finally, the last monopole to consider, $\mathfrak{M}_A^{-,-}$, can be dressed in a unique way:
\be
(\mathfrak{M}^{-,-}_A)_{B^{N_1-2}\,Q^{N_2-N_1}}\,\approx\,\mathfrak{M}_A^{-,-}\cdot \varepsilon^{(1)}_{a_1\,\dots a_{N_1-2}} \varepsilon_{(2)}^{b_1\,\dots\,b_{N_2-2}}\,B^{a_1}_{b_1}\cdots B^{a_{N_1-2}}_{b_{N_1-2}}\,Q^{i_1}_{b_{N_1-1}}\cdots Q^{i_{N_2-N_1}}_{b_{N_2-2}}\,.
\ee
This monopole always exists because $F\geq N_2-N_1$ by assumption.
\end{itemize}

As observed in \cite{Aharony:2013kma}, when studying the reduction of orthogonal gauge theories with a single node from four to three dimensions, a second type of baryonic operators play a relevant role. Such operators would correspond, in three dimensions, to {\it non-minimal} monopoles, {\it i.e.} monopoles with magnetic fluxes $m_1=-m_2=1$ for two different Cartan generators of the gauge group turned on. In $SO(N)$ SQCD with $F$ flavors, such monopoles are not actually chiral and do not map in a simple way across the duality. However, using the supersymmetric index, one can observe that monopoles with three magnetic fluxes turned on, that we will denote as $\mathfrak M^{\div,+}$ and $\mathfrak M^{+,\div}$, must be part of the chiral ring.\footnote{Observe that consistently with the result in $SO(N)$ SQCD, monopoles such as $\mathfrak{M}^{\div,0}$ and $\mathfrak{M}^{0,\div}$ are not chiral.}  The "symbol" "$\div$" is used to remember which node has two fluxes turned on; when such a monopole gets a VEV, the gauge group factor $SO(N_i)$ with two fluxes is broken down to $S(O(N_i-4)\times O(4))$: with the choice of fluxes $m_1=1=-m_2$, the monopole would not be invariant under the residual gauge group and it needs to be dressed with $N_i-4$ chiral fields contracted with the Levi-Civita symbol of the $SO(N_i-4)$ residual group:
\be
\begin{split}
&(\mathfrak{M}_A^{\div,+})_{B^{N_1-4}Q^{N_1-4}}\,\approx\,\mathfrak{M}^{\div,+}_A\cdot \varepsilon^{(1)}_{a_1\dots a_{N_1-4}} B^{a_1}_{b_1}Q^{b_1}_{i_1}\cdots B^{a_{N_1-4}}_{b_{N_1-4}}Q^{b_{N_1-4}}_{i_{N_1-4}}\,,\\
&(\mathfrak{M}_A^{+,\div})_{Q^{N_2-4}}\,\approx\, \mathfrak{M}^{+,\div}_A\cdot \varepsilon_{(2)}^{b_1\dots b_{N_2-4}}\, Q^{b_1}_{i_1}\cdots Q^{b_{N_2-4}}_{i_{N_2-4}}\,.
\end{split}
\ee
Their $R$-charges are:
\be
\begin{split}
&R[\mathfrak{M}_A^{\div,+}]\,=\,(2N_2+N_1-4)(1-R_B)+F(1-R_F)-(N_2-2)-(2N_1-6)\,,\\
&R[\mathfrak{M}_A^{+,\div}]\,=\,(2N_1+N_2-4)(1-R_B)+2F(1-R_F)-(N_1-2)-(2N_2-6)\,.
\end{split}
\ee

\subsubsection*{First dual}
Using the ARSW duality on the left orthogonal node of $\cT_A$ we obtain the following quiver theory:
\be
\scalebox{1}{
\begin{tikzpicture}[baseline]
\tikzstyle{every node}=[font=\footnotesize]
\node[draw=none] (node1) at (0,0) {$SO(N_2-N_1+2)$};
\node[draw=none] (node2) at (4,0) {$SO(N_2)$};
\node[draw, rectangle] (sqnode) at (7,0) {$F$};
\draw[black,solid] (node1) edge node[above]{$b$} (node2) ;
\draw[black,solid] (node2) edge node[above]{$Q$} (sqnode) ;
\draw[black,solid] (node2) edge [out=45,in=135,loop,looseness=5]node[above]{$S_B$} (node2) ;
\node[draw=none] at (4,-1) {$\mathcal W=\sigma_B \, \mathfrak M_B^{+, 0} + \Tr(bS_Bb)$};
\node[draw=none] at (-3,0) {$\mathcal T_{B}:$};
\end{tikzpicture}}\qquad \qquad
\ee
 where $S_B$ is a traceful symmetric field.  As usual, using a Seiberg-like duality $\text{Tr}(BB)$ maps to $\text{Tr}\,(S_B)$ and $Q_iQ_j$ maps to $Q_iQ_j$, implying the following constraints on the $R$-charges:
 \be
R_{S_B}\,=\,2(1-R_b)\,=\, 2R_B\,.
 \ee
The chiral ring generators consist of mesons, baryons and (baryonic) monopoles, built in the same fashion we presented for $\mathcal{T}_A$. Because of the constraints imposed by the superpotential, traces involving the bi-fundamental fields are set to zero and the chiral ring actually contains
\be
\text{Tr}(Q_iS_B^{J-1}Q_j)\,, \quad \text{Tr} (S_B^J)\,,\quad J=1,\dots,N_2\,.
\ee 
The theory again contains three baryons:
\be
\varepsilon_1 \varepsilon_2 b^{N_1} Q^{N_2-\tilde N_1}\,,\quad \varepsilon_1 b^{\tilde N_1} Q^{\tilde N_1}\,,\quad \varepsilon_2 Q^{N_2}\,,
\ee
where $\tilde N_1= N_2-N_1+2$. Observe that the bi-fundamental field $b$ appears in the definition of baryons since it is not traced but contracted with the Levi-Civita symbol. The monopole generators which are charge-conjugation even at the bare level are $\mathfrak M_B^{+,+}$ and $\left\{(\mathfrak M_B^{0,+})_{S_B^J}\right\}\,,J=0,\dots,N_2-2$, while $\mathfrak M_B^{+,0}$ is flipped by the singlet $\sigma_B$. Their R-charges are:
\be
\begin{split}
&R[\mathfrak{M}^{+,0}_B]\,=\,N_2(1-R_b)-(N_2-N_1)\,,\\
&R[\mathfrak{M}^{0,+}_B]\,=\,F(1-R_Q)\,+\,(N_2-N_1+2)(1-R_b)\,+\,N_2(1-R_S)-(N_2-2)\,,\\
&R[\mathfrak{M}^{+,+}_B]\,=\,F(1-R_Q)\,+\,(2N_2-N_1)(1-R_b)\,+\,N_2(1-R_S)-(2N_2-N_1-2)\,.
\end{split}
\ee
The baryonic monopoles (minimal and non-minimal) are defined as in the dual frame $\cT_A$ performing the substitution $\{B\,{\tiny \rightarrow}\,b,\,N_1\,{\tiny \rightarrow}\,\tilde N_1\}$. For this reason, we will not write down them again.

\subsubsection*{Second dual}

The last dual frame is obtained applying the ARSW duality to the $SO(N_2)$ node:
\be
\scalebox{1}{
\begin{tikzpicture}[baseline]
\tikzstyle{every node}=[font=\footnotesize]
%\node[draw, rectangle] (sqnodeL) at (-4,0) {$2$};
\node[draw=none] (node1) at (0,0) {$SO(N_1)$};
\draw[black,solid] (node1) edge [out=225,in=135,loop,looseness=5]node[left]{$S_C$} (node1) ;
\node[draw=none] (node2) at (4,0) {$SO\left(F+N_1-N_2+2\right)$};
\node[draw, rectangle] (sqnode) at (2,2.5) {$F$};
\draw[black,solid] (sqnode) edge [out=45,in=135,loop,looseness=5]node[above]{$M$} (sqnode) ;
\draw[black,solid] (node1) edge node[below]{$\tilde b$} (node2) ;
\draw[black,solid] (node2) edge node[above]{$\tilde q$} (sqnode) ;
\draw[black,solid] (node1) edge node[above]{$p$} (sqnode) ;
\node[draw=none] at (2.2,-1) {$\mathcal W=\sigma_C \, \mathfrak M_C^{0,+} + \Tr( \tilde bS_C \tilde b) + \Tr( \tilde bp \tilde q)+\Tr(\tilde qM \tilde q)$};
\node[draw=none] at (-1,1.5) {$\mathcal T_{C}:$};
\end{tikzpicture}}
\ee
The usual map of the ARSW duality implies $\text{Tr}(BB)\rightarrow \text{Tr}S_C$ and $Q_iQ_j\rightarrow M_{ij}$ fixing the chiral field R-charges in the third frame:
\be
R_{S_C}=2(1-R_b)=2R_B\,,\quad R_{M}=2(1-R_q)=2R_Q\,,\quad R_p\,=\,R_B+R_Q\,.
\ee
Given the constraints imposed by the superpotential, the mesonic operators are ${\Tr}(p_iS_C^Jp_j)$ and ${\Tr}(S_C^J)$, while the conjugation-even (dressed) monopoles are $\mathfrak{M}^{+,+}$ and $\left\{(\mathfrak M^{+,0})_{S_C^J}\right\}$; the monopole $\mathfrak{M}^{0,+}$ is flipped by $\sigma_C$ instead. Again, as in the other frames, there are three different baryons: 
\be
\label{eq:baryonTCSOSO}
\varepsilon_1 \tilde b^{N_1}\tilde q^{\tilde N_2-N_1}\,,\quad \varepsilon_1 p^{N_1}\,,\quad \varepsilon_2 \tilde q^{\tilde N_2}\,,
\ee
where $\tilde N_2= N_1+F-N_2+2$. Finally we have to define the baryon monopoles in this duality frame. This time, they need some particular attention because there is no simple rules to relate them to the ones in $\cT_A$. There are two monopoles of charge $(-,+)$ with respect to $(\cC_1,\cC_2)$ that are dressed using the chiral field $p$ only:
\be
(\mathfrak{M}^{-,0}_C)_{p^{N_1-2}}\,,\quad (\mathfrak{M}^{-,+}_C)_{p^{N_1-2}}\,,
\ee
while there are two baryon monopoles of charge $(+,-)$ that are dressed using $\tilde q$:
\be
(\mathfrak{M}^{0,-}_C)_{\tilde q^{\tilde N_2-2}}\,,\quad (\mathfrak{M}^{+,-}_C)_{\tilde q^{\tilde N_2-2}}\,.
\ee
Finally, there are three monopoles constructed using $\tilde b$ and $\tilde q$ appropriately contracted with Levi-Civita symbols:
\be
(\mathfrak{M}^{-,-}_C)_{\tilde b^{N_1-2}\,\tilde q^{\tilde N_2-N_1}}\,, \quad (\varepsilon_2\mathfrak{M}^{-,0}_C)_{\tilde b^{N_1-2}\tilde q^{\tilde N_2-N_1+2}}\,, \quad  ({\varepsilon_{1}\mathfrak{M}}^{0,-}_C)_{\,\tilde b^{N_1}\,\tilde q^{\tilde N_2-N_1-2}}\,.
\ee
Observe, that the last two operators in the previous equation are the dibaryon monopole of $\cT_C$. Finally, the non-minimal monopoles are $(\mathfrak{M}_C^{\div,+})_{p^{N_1-4}}$ and $(\mathfrak{M}_C^{+,\div})_{\tilde q^{\tilde N_2-4}}$. For definiteness, let us write the $R$-charges of the minimal fluxes in absence of any additional dressing:
\be
\begin{split}
&R[\mathfrak{M}^{+,0}_C]\,=\,(F+N_1-N_2+2)(1-R_{\tilde{b}})+N_1(1-R_S)+F(1-R_p)-(N_1-2)\,,\\
&R[\mathfrak{M}^{0,+}_C]\,=\, N_1(1-R_{\bt})+F(1-R_{\qt})-(F+N_1-N_2)\,,\\
&R[\mathfrak{M}^{+,+}_C]\,=\, (F+2N_1-N_2)(1-R_{\tilde{b}})+F(2-R_p-R_{\qt})+N_1(1-R_S)-(F+2N_1-N_2-2)\,.
\end{split}
\ee

\subsubsection*{Operator map}
The operator map across the triality is quite intricate:
\be
\label{OMSOSO}
\begin{tabular}{c c c c c}
$\mathcal T_A$ & & $\mathcal T_B$    & &  $\mathcal T_C$\\
\hline
$\mathfrak{M}_A^{+,0}$ & & $\sigma_B$    & &  $\mathfrak{M}_C^{+,+}$\\
$\mathfrak{M}_A^{0,+}$ & & $\mathfrak{M}_B^{+,+}$    & &  $\sigma_C$\\
$(\mathfrak{M}_A^{+,+})_{(B\!B)^J}$ & & $(\mathfrak{M}_B^{0,+})_{S_B^J}$    & &  $(\mathfrak{M}_C^{+,0})_{S_C^J}$\\
\hline
$(\mathfrak{M}^{-,-}_A)_{B^{N_1-2}\,Q^{N_2-N_1}}$ & & $(\mathfrak{M}^{-,+}_B)_{b^{\tilde N_1-2}\,Q^{\tilde N_1-2}}$    & &  $(\mathfrak{M}^{+,-}_C)_{\tilde q^{\tilde N_2-2}}$\\
$(\mathfrak{M}^{+,-}_A)_{Q^{N_2-2}}$ & &  $(\mathfrak{M}^{0,-}_B)_{Q^{N_2-2}}$   & &  $(\mathfrak{M}^{-,-}_C)_{\tilde b^{N_1-2}\,\tilde q^{\tilde N_2-N_1}}$\\
$(\mathfrak{M}^{-,+}_A)_{B^{N_1-2}\,Q^{N_1-2}}$  & & $(\mathfrak{M}^{-,-}_B)_{b^{\tilde N_1-2}\,Q^{N_2-\tilde N_1}}$    & &  $(\mathfrak{M}^{-,0})_{p^{N_1-2}}$\\
$(\mathfrak{M}^{-,0}_A)_{B^{N_1-2}\,Q^{N_1-2}}$  & & $\varepsilon_1\varepsilon_2b^{\tilde N_1}Q^{N_2-\tilde N_1}$    & &  $(\mathfrak{M}^{-,+}_C)_{p^{N_1-2}}$\\
$(\varepsilon_2\mathfrak{M}^{-,0}_A)_{B^{N_1-2}Q^{N_2-N_1+2}}$  & & $\varepsilon_1b^{\tilde N_1}Q^{\tilde N_1}$    & &  $(\mathfrak{M}_C^{+,\div})_{\tilde q^{\tilde N_2-4}}$\\
$(\mathfrak{M}^{0,-}_A)_{Q^{N_2-2}}$  & & $(\mathfrak{M}^{+,-}_B)_{Q^{N_2-2}}$    & &  $\varepsilon_1 \tilde b^{N_1}\tilde q^{\tilde N_2-N_1}$\\
$({\varepsilon_{1}\mathfrak{M}}^{0,-}_A)_{\,B^{N_1}\,Q^{N_2-N_1-2}}$  & & $(\mathfrak{M}_B^{\div,+})_{b^{\tilde N_1-4}Q^{\tilde N_1-4}}$    & &  $\varepsilon_2 \tilde q^{\tilde N_2}$\\
\hline
$(\mathfrak{M}_A^{\div,+})_{B^{N_1-4}Q^{N_1-4}}$ &&  $({\varepsilon_{1}\mathfrak{M}}^{0,-}_B)_{\,b^{\tilde N_1}\,Q^{N_2-\tilde N_1-2}}$    & &  $(\mathfrak{M}_C^{\div,+})_{p^{N_1-4}}$\\
$(\mathfrak{M}_A^{+,\div})_{Q^{N_2-4}}$ &&  $(\mathfrak{M}_B^{+,\div})_{Q^{N_2-4}}$    & &  $(\varepsilon_2\mathfrak{M}^{-,0}_C)_{\tilde b^{N_1-2} \tilde q^{\tilde N_2-N_1+2}}$\\
\hline
$\varepsilon_1\varepsilon_2B^{N_1}Q^{N_2-N_1}$    &&    $(\mathfrak{M}^{-,0}_B)_{b^{\tilde N_1-2}\,Q^{\tilde N_1-2}}$    & &  $(\mathfrak{M}^{0,-}_C)_{\tilde q^{\tilde N_2-2}}$\\
$\varepsilon_1B^{N_1}Q^{N_1}$    &&    $(\varepsilon_2\mathfrak{M}^{-,0}_B)_{b^{\tilde N_1-2}Q^{N_2-\tilde N_1+2}}$    & &  $\varepsilon_1 p^{N_1}$\\
$\varepsilon_2{Q}^{N_2}$    &&    $\varepsilon_2{Q}^{N_2}$    & &  $({\varepsilon_{1}\mathfrak{M}}^{0,-}_C)_{\,\tilde b^{N_1}\,\tilde q^{\tilde N_2-N_1-2}}$\\
\hline
$Q_iQ_j$   &&     $Q_iQ_j$    & &  $M_{ij}$\\
$\text{Tr}((BB)^J)$   &&     $\text{Tr}(S_B^J)$    & &  $\text{Tr}(S_C^J)$\\
$\text{Tr}(Q_i(B\!B)^JQ_j)$   &&     $\text{Tr}(Q_i\,S_B^{J}\,Q_j)$    & &  $\text{Tr}(p_i\,S_C^{J-1}\,p_j)$
\end{tabular}
\ee
Let us observe that the presence of the non-minimal monopole is crucial in order to have a consistent map among the three frames. It is interesting to observe that, in order the previous map to be consistent, the discrete charge-conjugation symmetries must be also identified as follows across the duality:
\be
\begin{tabular}{c c c c c}
$\cC_1^A$   & $\leftrightarrow$  & $\cC_1^B$ &  $\leftrightarrow$ & $\cC_1^C\cdot \cC_2^C$ \\
$\cC_2^A$   & $\leftrightarrow$   & $\cC_1^B\cdot \cC_2^B$ &  $\leftrightarrow$  & $\cC_2^C\,.$
\end{tabular}
\ee

\subsubsection*{Supersymmetric index}

In this section we present the supersymmetric index of the three dual theories with the particular choice case of $N_1=3$, $N_2=4$ and $F=3$. We fix the $R$-charges to be $R_B=\frac{1}{2}$ and $R_Q=\frac{1}{3}$; moreover, the charges under the $U(1)$ global symmetries are fixed as in \eqref{eq:chU(1)SpSO}. As before, we denote with $y_B$ and $y_Q$ the fugacities of $U(1)_B$ and $U(1)_Q$ respectively, while we denote with $x$ the $R$-charge fugacity and with $z_i$ the $SU(F)$ ones.The supersymmetric index reads:
\be
\begin{split}
&\mathcal{I}\,=\,\\
&1+{\color{blue}6\,x^{2/3}\,y_Q^2}\,+\,{\color{blue}x\left(y_B^2+\frac{1}{y_B^4}\right)}\,+\,21\,x^{4/3}\,y_Q^4\,+\,{\color{blue}x^{3/2}\left( \frac{1}{y_Q^3\,y_B^5}+\frac{1}{y_Q^3\,y_B^3} \right)}+x^{5/3}\left( \frac{6\,y_Q^2}{y_B^4}+\,{\color{blue}12\,y_Q^2\,y_B^2} \right)+\\
&+{\color{blue}x^{11/6}\left( \frac{3\,y_Q}{y_B^3}+3\,y_Q\,y_B^3 \right)}+x^2\left(\frac{1}{y_B^8}+\frac{1}{y_B^2}+2\,y_B^2+56\,y_Q^6-10 \right)+{\color{blue}x^{13/6}\,\left( \frac{9}{y_Q\,y_B^5}+\frac{9}{y_Q\,y_B^3} \right)}\,+\,\\
&x^{7/3}\left( {\color{blue}\frac{6}{y_Q^2\,y_B^2}}+\frac{21\,y_Q^4}{y_B^4}+60\,y_Q^4\,y_B^2 \right)\,+\,x^{5/2}\left(\frac{1}{y_Q^3\,y_B^9}+\frac{2}{y_Q^3\,y_B^3}+{\color{blue}\frac{19\,y_Q^3}{y_B^3}}+\frac{1}{y_Q^3\,y_B}+{\color{blue}19\,y_Q^3\,y_B^3}\right)\,+\,\dots
\end{split}
\ee
We highlighted in blue the generators of the chiral ring and correspond to:
\begin{align}
\nonumber
	\Tr(Q_iQ_j) \;&\leftrightarrow\; {\color{blue}6\,x^{2/3}\,y_Q^2} & \mathfrak{M}_A^{+,0} \;&\leftrightarrow\; {\color{blue}\frac{x}{y_B^4}} \\
\nonumber
	\text{Tr}(BB) \;&\leftrightarrow\; {\color{blue}x\,y_B^2}  &\mathfrak{M}_A^{+,+} \; &\leftrightarrow\; {\color{blue}\frac{x^{3/2}}{y_Q^3\,y_B^5}} \\
\nonumber
	\mathfrak{M}_A^{0,+} \;&\leftrightarrow\; {\color{blue}\frac{x^{3/2}}{y_Q^3\,y_B^3}} & \text{Tr}(Q_i(BB)Q_j) \;&\leftrightarrow\; {\color{blue}6\,x^{5/3}\,y_Q^2\,y_B^2} \\
\nonumber	
\varepsilon_1\varepsilon_2B^{N_1}Q^{N_2-N_1} \;&\leftrightarrow\; {\color{blue}3\,x^{11/6}\,y_Q\,y_B^3} &(\mathfrak{M}^{-,0}_A)_{B^{N_1-2}\,Q^{N_1-2}} \;&\leftrightarrow\; {\color{blue}\frac{3\,x^{11/6}\,y_Q}{y_B^3}} \\
\nonumber	
	\text{Tr}(BB)^2 \;&\leftrightarrow\; {\color{blue}x^2\,y_B^4} &(\mathfrak{M}^{+,-}_A)_{Q^{N_2-2}}\;&\leftrightarrow\; {\color{blue}\frac{3\,x^{13/6}}{y_Q\,y_B^5}} \\
\nonumber	
(\mathfrak{M}^{0,-}_A)_{Q^{N_2-2}} \;&\leftrightarrow\; {\color{blue}\frac{3\,x^{13/6}}{y_Q\,y_B^3}} & (\mathfrak{M}^{-,+}_A)_{B^{N_1-2}\,Q^{N_1-2}} \;&\leftrightarrow\; {\color{blue}\frac{3\,x^{7/3}}{y^2_Q\,y_B^4}} \\
\nonumber	
(\varepsilon_2\mathfrak{M}^{-,0}_A)_{B^{N_1-2}Q^{N_2-N_1+2}}\;&\leftrightarrow\; {\color{blue}\frac{3\,x^{7/3}}{y^2_Q\,y_B^4}} & (\mathfrak{M}_A^{+,+})_{(BB)} \;&\leftrightarrow\; {\color{blue}\frac{2\,x^{5/2}}{y^3_Q\,y_B^3}} \\
\nonumber	
(\varepsilon_2\mathfrak{M}^{-,0}_A)_{B^{N_1-2}Q^{N_2-N_1+2}} \;&\leftrightarrow\; {\color{blue}\frac{x^{5/2\,y_Q^3}}{\,y_B^3}} &\varepsilon_1B^{N_1}Q^{N_1} \;&\leftrightarrow\; {\color{blue}x^{5/2}\,y^3_Q\,y_B^3} \\
\end{align}
Let us observe that, for this choice of ranks of the gauge groups, not all the possible baryon and baryon monopoles are present. In particular, there are no non-minimal monopoles for $\cT_A$ with the current choice.

%%%%%%%%%%%%%%%%%%%%%%%%%%%%%%%%%%%%%%%%%%%%%%%%%%%%%%%%%%%%%%%
%%%%%%%%%%%%%%%%%%%%%%%%%%%%%%%%%%%%%%%%%%%%%%%%%%%%%%%%%%%%%%%
%%%%%%%%%%%%%%%%%%%%%%%%%%%%%%%%%%%%%%%%%%%%%%%%%%%%%%%%%%%%%%%

\subsection{Mapping the orthogonal baryonic operators in general quivers}
Using the operator maps  \eqref{OMSOSp} and \eqref{OMSOSO} we can guess how baryons and baryon monopoles are mapped across the duality in more general quivers. In the case of ortho-symplectic quivers, we propose precise mapping rules based on the two-node experience. We propose that, applying an ARSW duality to the $i^{th}$ node:
\begin{itemize}
\item if the node is symplectic, the general rule presented in the introduction still applies. For instance:
\be
(\mathfrak{M}^{\dots,\sigma_{i-1},\bullet_i,\sigma_{i+1},\dots})\rightarrow (\mathfrak{M}^{\dots,\sigma_{i-1},\bullet_i,\sigma_{i+1},\dots})\,,\quad  (\mathfrak{M}^{\dots,\sigma_{i-1},\bullet_i,0_{i+1},\dots})\rightarrow (\mathfrak{M}^{\dots,\sigma_{i-1},0_i,0_{i+1},\dots})
\ee
and so on, where $\sigma_i=\pm 1$.

\item If the node is special-orthogonal and the monopole is conjugation-even with respect to $\cC_i$, the flux is turned on or turned off following the same rules as the symplectic nodes. For instance $(\mathfrak M^{\dots,-_{i-1},0_i,0_{i+1},\dots})\rightarrow (\mathfrak M^{\dots,-_{i-1},+_i,0_{i+1},\dots})$ and so on.

\item If the node is special-orthogonal and the monopole is of the type $(\mathfrak{M}^{0,\dots0_{i-1}, -_i,0_{i+1},\dots,0})$, it maps to a baryon of the type $\varepsilon_i\cdots$.

\item  If the node is special-orthogonal and the monopole is of baryonic type with respect to $\cC_i$, a monopole $(\mathfrak M^{\dots,\omega_{i-1}, -_i, \omega_{i+1},\dots})$ with at least one between $\omega_{i+1}$ and $\omega_{i-1}$ non vanishing, the operator is mapped to another baryon monopole with the same string of fluxes.
\end{itemize}
Observe that in the case the monopole is baryonic, we do not indicate how the dressing transforms across the duality, such task can be easily addressed matching the quantum numbers between the frames. The knowledge of the mapped string of fluxes is the crucial guideline in order to do so.

When the quiver is purely special-orthogonal, instead, the situation is much more intricate and we do not really have a general prescription but we conjecture some general guideline to follow and that should simplify the task of matching the chiral ring operators. The higher intricacy is mostly due to the fact that we do not know in general which strings of fluxes are admitted, in particular for the monopoles of non-minimal or dybarion type. Let us consider a monopole of the form $\mathfrak M^{\dots\,\omega_{i-1}\,\omega_{i}\,\omega_{i+1}\,\dots}$ where $\omega_j=0,1,-1$ denotes the bare charges of the monopole with respect to the charge conjugation symmetry of the $j$-th node; the operators will be assumed to be not dibaryons or non-minimal unless explicitly stated. Let us assume to apply an ARSW duality to the $i^{th}$ node. We conjecture that:
\begin{itemize}
\item if $\sigma_i=+,0$ the monopole keeps on transforming as explained in the introduction for quivers without baryons. For instance, the monopole $\mathfrak{M}^{\cdots\,0,+,\omega_{i+1}\,\cdots}$ maps to a monopole of the form $\mathfrak{M}^{\cdots\,0,0,\omega_{i+1},\cdots}$.

\item Monopoles of the form $\mathfrak M^{0,\cdots,0_{i-1},-_i,0_{i+1},\cdots,0}$ map to baryons.

\item If $\omega_i=-1$ and at least one between $\omega_{i\pm1}$ differs from zero, $\mathcal{M}^{\cdots, \omega_{i-1}, -, \omega_{i+1},\cdots}$ maps to a monopole with $\omega_{i\pm 1}\rightarrow-\omega_{i\pm 1}$.

\item We conjecture that a non-minimal monopole to be of the form $(\mathfrak{M}^{0,\dots,{0_{i+1},\div_i,+_{i-1}},\dots})$ and it gets mapped to a dibaryon monopole of the form $(\varepsilon_i\mathfrak{M}^{0,\dots,{0_{i+1},0_i,-_{i-1}},\dots})$ and viceversa.
\end{itemize}
It would be interesting in the future to further investigate the precise content of the chiral ring in longer orthogonal quivers in order to make the previous conjectured prescription more precise.

%%%%%%%%%%%%%%%%%%%%%%%%%%%%%%%%%%%%%%%%%%%%%%%%%%%%%%%%%%%%%%%
\subsection{$Sp (N_1) \times SU(N_2)$ theory}\label{SPSU}
In the spirit of studying dualities in two-node quivers, let us consider a theory with $Sp(N_1)\times
SU(2N_2)$ gauge group:
\be
\scalebox{1}{
\begin{tikzpicture}[baseline]
\tikzstyle{every node}=[font=\footnotesize]
%\node[draw, rectangle] (sqnodeL) at (-4,0) {$2$};
\node[draw=none] (node1) at (0,0) {$Sp(N_1)$};
\node[draw=none] (node2) at (2.5,0) {$SU(2N_2)$};
\node[draw, rectangle] (sqnode1) at (5,0) {$2F$};
\node[draw, rectangle] (sqnode2) at (2.5,-2) {$2F+2N_1$};
\draw[black,solid,<-] (node1) edge node[above]{$B$} (node2) ;
\draw[black,solid,->] (node2) edge node[above]{$Q_1$} (sqnode1) ;
\draw[black,solid,<-] (node2) edge node[right]{$Q_2$} (sqnode2) ;
%\draw[black,solid,->] (node2) edge [out=25,in=155,loop,looseness=1]node[above]{$F$} (sqnodeR) ;
\node[draw=none] at (1,-1) {$\mathcal W=0$};
\node[draw=none] at (-2,0) {$\mathcal T_{A}:$};
%node[above]{\blue $X$} node[below, yshift=0cm] {$\red F_X$} node {$ \red \times$}  (node);
\end{tikzpicture}}
\ee
where the $SU(2N_2)$ gauge node has the same number of fundamental and anti-fundamental chirals, hence it is non-chiral. The global symmetry is $SU(2F) \times SU(2F+2N_1) \times U(1)_B\times U(1)_{Q_1}\times U(1)_{Q_2}$, where the three $U(1)$'s act on the three sets of chirals $B$, $Q_1$ and $Q_2$ as indicated by the corresponding label.

The $R$-charge of monopole operators $\mathfrak M^{\vec m, \vec n}$, $\vec m$ being the magnetic charges for the $Sp(N_1)$ node and $\vec n$ the ones for $SU(2N_2)$\footnote{Recall that the minimal monopole for an $SU(N)$ gauge theory has magnetic charges $\vec m=(1,0^{N-2},-1)$ breaking the gauge group down to $SU(N-2)\times U(1) \times U(1)$.} 
\be
\begin{split}
	R[\mathfrak M_A^{\vec m, \vec n}]&=\frac{1}{2}(1-r_B) \sum_{a=1}^{N_1} \sum_{\sigma=0,1}
	\sum_{b=1}^{2N_2} |(-1)^\sigma m_a - n_b| + \frac{1}{2}2F(1-r_{Q_1})						\sum_{b=1}^{2N_2} |n_b|
	\\
	&+ \frac{1}{2}(2F+2N_1)(1-r_{Q_2}) \sum_{b=1}^{2N_2} |n_b|
	-\sum_{b_1<b_2} |n_{b_1} -n_{b_2}| \\
	&-\sum_{a_1<a_2} (|m_{a_1}-m_{a_2}|+|m_{a_1}+m_{a_2}|)-\sum_{a=1}^{N_1} |2m_a|
\end{split}
\ee
In particular, the monopoles with minimal GNO fluxes have $R$-charges
\be
	R[\mathfrak M_A^{\cdot, 0}] = 2N_2(1-r_B) - 2N_1,
\ee
\be
	R[\mathfrak M_A^{0, \cdot}] =2N_1(1-r_B) + 2F (1-r_{Q_1}) 
	+(2F+2N_1) (1-r_{Q_2}) - 2(2N_2-1)
\ee
\be
\begin{split}
	R[\mathfrak M_A^{\cdot, \cdot}]&=2F (1-r_{Q_1})+(2F+2N_1) (1-r_{Q_2}) \\
	&+(2N_1+2N_2-2)(1-r_B)-2N_1 -2(2N_2-1)
\end{split}
\ee
The monopole with flux both on the $Sp(N_1)$ and on $SU(N_1)$ deserves particular attention and will be discussed at length in the following.
The chiral ring of this theory has generators including mesonic and baryonic operators and dressed monopoles. We can construct a tower of baryonic operators with an even number of bifundamental fields $B$ and the appropriate number of $Q_1$ in order to use the $SU(2N_2)$ epsilon tensor
\be
	B^{i_1}_{a_1} \dots B^{i_{2k}}_{a_{2k}} {Q_1}^{i_{2k+1}} \dots {Q_1}^{i_{2N_2}}
	\Omega_{a_1 a_2} \dots \Omega_{a_{2k-1} a_{2k}} \epsilon_{i_1 \dots i_{2N_2}},
\ee
where the $i$ indices are the ones for the $SU(2N_2)$ gauge group, the $a$ are for the $Sp(N_1)$ and are contracted using the symplectic form $\Omega$. For simplicity we suppressed the indices of the flavour symmetry.
This operator transforms in the 
rank-$(2N_2-2k)$ antisymmetric representation of the $SU(2F)$ global symmetry, the antisymmetry being induced by the contraction with the epsilon tensor.
Furthermore, we may construct a mesonic-like operator using $B$ and $Q_2$ as follows
\be
	B^{i_1}_{a_1} B^{i_2}_{a_2} (Q_2)_{i_3} (Q_2)_{i_4} \delta^{i_3}_{i_1} \delta^{i_4}_{i_2}
	\Omega_{a_1 a_2}.
\ee 
Finally we have the meson $\tr(Q_1 Q_2)$.

\subsubsection*{First dual}
Let us apply Aharony duality to $Sp(N_1)$. We obtain the following dual theory 
\be
\scalebox{1}{
\begin{tikzpicture}[baseline]
\tikzstyle{every node}=[font=\footnotesize]
%\node[draw, rectangle] (sqnodeL) at (-4,0) {$2$};
\node[draw=none] (node1) at (0,0) {$Sp(N_2-N_1-1)$};
\node[draw=none] (node2) at (3.5,0) {$SU(2N_2)$};
\node[draw, rectangle] (sqnode1) at (6,0) {$2F$};
\node[draw, rectangle] (sqnode2) at (3.5,-2) {$2F+2N_1$};
\draw[black,solid,->] (node1) edge node[above]{$b$} (node2) ;
\draw[black,solid,->] (node2) edge node[above]{$Q_1$} (sqnode1) ;
\draw[black,solid,<-] (node2) edge node[right]{$Q_2$} (sqnode2) ;
\draw[black,solid] (node2) edge [out=45,in=135,loop,looseness=5]node[above]{$A$} (node2) ;
%\draw[black,solid,->] (node2) edge [out=25,in=155,loop,looseness=1]node[above]{$F$} (sqnodeR) ;
\node[draw=none] at (1,-1) {$\mathcal W=\sigma_B \mathfrak M_B^{\cdot, 0} + \Tr(b A b)$};
\node[draw=none] at (-2.5,0) {$\mathcal T_{B}:$};
%node[above]{\blue $X$} node[below, yshift=0cm] {$\red F_X$} node {$ \red \times$}  (node);
\end{tikzpicture}}
\ee
where $A$ is an antisymmetric field. Standard Aharony duality would imply the map $\Tr(B B) \leftrightarrow A$; however, being the flavour symmetry of the $Sp(N_1)$ gauged in this case, $A$ is not a gauge invariant operator, so we have to look at 
$\Tr(B B Q_1 Q_1) \leftrightarrow \Tr(A Q_1 Q_1)$, implying $R_A=2R_B$. From the superpotential we also get $R_b=1-R_B$.
The $R$-charge of monopole operator $\mathfrak M^{\vec m, \vec n}$, reads
\be
\begin{split}
	R[\mathfrak M_B^{\vec m, \vec n}]&=\frac{1}{2}(1-R_b) \sum_{a=1}^{N_2-N_1-1} \sum_{\sigma=0,1}
	\sum_{b=1}^{2N_2} |(-1)^\sigma m_a - n_b| + \frac{1}{2}2F(1-R_{Q_1})						\sum_{b=1}^{2N_2} |n_b|
	\\
	&+ \frac{1}{2}(2F+2N_1)(1-R_{Q_2}) \sum_{b=1}^{2N_2} |n_b|+ \frac{1}{2} (1-R_A) \sum_{b_1<b_2} |n_{b_1}+n_{b_2}|\\
	&-\sum_{b_1<b_2} |n_{b_1} -n_{b_2}| -\sum_{a_1<a_2} (|m_{a_1}-m_{a_2}|+|m_{a_1}+m_{a_2}|)-\sum_{a=1}^{N_2-N_1-1} |2m_a|.
\end{split}
\ee
For the minimal GNO flux monopoles we find
\begin{align}
	R[\mathfrak M_B^{\cdot, 0}]&=2N_2(1-R_b)-2(N_2-N_1-1),\\ 
	R[\mathfrak M_B^{0, \cdot}]&=(2N_2-2N_1-2)(1-R_b)+2F(1-R_{Q_1}) 
	+(2F+2N_1)(1-R_{Q_2}) \notag \\
	&+(2N_2-2)(1-R_A)-2(2N_2-1), \\
	R[\mathfrak M_B^{\cdot, \cdot}]&=(4N_2-2N_1-4)(1-R_b)+2F(1-R_{Q_1})
	+(2F+2N_1)(1-R_{Q_2}) \notag\\
	&+(2N_2-2)(1-R_A)-2(N_2-N_1-1)-2(2N_2-1),
\end{align}
This theory has interesting dressed monopoles generating the chiral ring that needs to be discussed. The interesting part of the discussion comes from the presence of the antisymmetric field $A$ for the $SU(2N_2)$ group. 

\paragraph{Dressed monopoles in a $SU(2N)$ theory with an antisymmetric field}
The $SU(2N)$ theory with an antisymmetric field, and in 
particular its dressed monopoles, has been discussed in \cite{Nii:2019ebv}. We will closely follow this reference to review the construction of the dressed monopoles. For the moment we generalize the set-up of \cite{Nii:2019ebv} by taking into account an $SU(2N)$ gauge theory with $N_f$ fundamentals $Q$, ${\bar N}_f$ anti-fundamentals, $N_A$ antisymmetric and ${\bar N}_A$ conjugate antisymmetric. As already observed, a monopole with minimal GNO flux for $SU(2N)$ breaks the gauge group to $SU(2N-2) \times U(1)_1\times U(1)_2$, where $1$ and $2$ attached to the $U(1)$'s are just labels. It turns out that there is a mixed CS term between these two $U(1)$'s:
\be\label{keff}
	k_{\text{eff}}^{U(1)_1,U(1)_2}=(2N-2)(N-2)(N_A-\bar N_A) + (N-1)(N_f-\bar N_f).
\ee
This mixed CS term induces a gauge charge under $U(1)_2$ for the bare monopole \cite{Intriligator:2013lca} given by 
\be
	U(1)_2[\mathfrak M]=-k_{\text{eff}}^{U(1)_1,U(1)_2}.
\ee
Therefore, to construct gauge invariant monopoles we need to consider a dressing. To this end, we have to look at the decomposition of the matter fields under the residual gauge group $SU(2N-2) \times U(1)_1\times U(1)_2$:
%\begin{align}
%	&[0,1,0^{2N-3}] \to [0,1,0^{2N-5}]_{0,-2} +  [1,0^{2N-4}]_{1,N-2} + 				[1,0^{2N-4}]_{-1,N-2} + [0^{2N-3}]_{0,2(N-1)} \notag\\
%	&[0^{2N-3},1,0] \to [0^{2N-5},1,0]_{0,-2} +  [0^{2N-4},1]_{-1,-(N-2)} + 				[0^{2N-4},1]_{1,-(N-2)} + [0^{2N-3}]_{0,-2(N-1)} \notag\\
%	&[1,0^{2N-2}] \to [1,0^{2N-4}]_{0,-1} + [0^{2N-3}]_{1,N-1} + [0^{2N-3}]_{-1,N-1}
%	\notag\\
%	&[0^{2N-2},1] \to [0^{2N-4},1]_{0,1} + [0^{2N-3}]_{-1,-(N-1)} + [0^{2N-3}]_{1,-(N-1)} 
%\end{align}
%
\begin{align}
	&\text{asym} \to \text{asym}_{0,-2} +  \text{fund}_{1,N-2} + \text{fund}_{-1,N-2} + 
	\text{sing}_{0,2(N-1)} \notag\\
	&\overbar{\text{asym}} \to \overbar{\text{asym}} _{0,-2} +  \overbar{\text{fund}} _{-1,-(N-2)} + 	\overbar{\text{fund}} _{1,-(N-2)} + {\text{sing}} _{0,-2(N-1)} \notag\\
	&{\text{fund}} \to {\text{fund}}_{0,-1} + {\text{sing}}_{1,N-1} + {\text{sing}}_{-1,N-1}
	\notag\\
	&\overbar{\text{fund}}  \to \overbar{\text{fund}} _{0,1} + {\text{sing}} _{-1,-(N-1)} + {\text{sing}} _{1,-(N-1)} 
\end{align}
Now we may adapt the set-up to our theory $\mathcal T_B$, hence we take $N_A=1, \,\bar N_A=0, \, N_f=2F, \, \bar N_f=2F+2N_2-2$. The gauge charge of the bare monopole is now given by
\be
	U(1)_2[\mathfrak M]=2N-2.
\ee
The crucial point is now to use matter fields in the residual gauge theory that may be used to cancel this gauge charge. The residual antisymmetric field $\text{asym}_{0,-2}$ is a good candidate to do this job. However we need to take $N-1$ copies of it, so the $U(1)_2$ charge is cancelled and there is no $U(1)_1$ charge brought by the dressing. In the end, the gauge invariant monopole has the form
\be
	\{ \mathfrak M_{A^{N-1}} \},
\ee
where $A^{N-1}$ is contracted using the epsilon tensor of the residual $SU(2N-2)$
\be
	A^{i_1 i_2} \dots A^{i_{2N-1} i_{2N}} \epsilon_{i_1 \cdots i_{2N}}.
\ee
This construction may be generalised to get a tower of dressed monopoles. The fundamental field in the residual theory can be used in a  way similar to the antisymmetric. Since the $U(1)_2$ charge is $-1$ and not $-2$ as the antisymmetric, every time we ``remove" one $A$ from $\{ \mathfrak M_{A^{N-1}} \}$ we put two fundamentals $Q$'s. In the end, the tower of dressed monopoles that we get has the form
\be
	\{ \mathfrak M_{A^{N-1-k} Q^{2k}} \} \qquad \text{for} \; k=0,\dots, N-1.
\ee

We can now go back to our quiver gauge theory. The presence of a tower of dressed monopoles reflects the fact that the chiral ring of $\mathcal T_B$ includes the monopoles $\{(\mathfrak M_B^{0,\cdot})_{A^{N_2-1-k} Q_1^{2k}} \}$. Furthermore, the chiral ring includes the singlet $\sigma_B$ and the monopole with flux on both gauge nodes $\mathfrak M_B^{\cdot, \cdot}$. The mesonic part of the chiral ring generators is given by the mesons $\tr(Q_1 Q_2)$ and $\tr(A Q_2 Q_2)$. One may also construct baryonic operators. First, we have $\varepsilon Q_2^{2N_2}$, but we can also form baryons using the antisymmetric $A$ and the fundamental chirals $Q_1$. In detail, we get a tower of baryons:
\be
	A^{i_1 i_2}\cdots A^{i_{2k-1} i_{2k}} Q_1^{i_{2k+1}} \cdots Q_1^{i_{2N_2}} \epsilon_{i_1 \dots i_{2N_2}},
\ee
where again we suppressed the flavor indices of $Q_1$. Again, due to the anti-symmetrization, the baryonic operators that we denote schematically as $\varepsilon A^k Q_1^{2N_2-2k}$ transform in the rank-$(2N_2-k)$ antisymmetric representation of the $SU(2F)$.

Let us also comment on the monopole with both fluxes on the $Sp(N_2-N_1-1)$ and on the 
$SU(2N_2)$. In this case also the symplectic gauge node is broken down as $Sp(N_2-N_1-1) \to Sp(N_2-N_1-2) \times SU(2).$ Hence, if we want to compute the $U(1)_2$ charge of the monopole $\mathfrak M^{\cdot, \cdot}$ we need to take into account that the contribution from the bifundamental fields $b$ is now reduced to $2(N_2-N_1-2)$. In detail, applying \eqref{keff} with $N_A=1, \, \bar N_A=0, \, N_f=2F, \,
\bar N_f=(2F+2N_1) + (2N_2-2N_1-4)$ we find $U(1)_2[\mathfrak {M}^{\cdot, \cdot}]=0$.
So the monopole $\mathfrak {M}^{\cdot, \cdot}$ is gauge invariant, and is a generator of the chiral ring

%\be
%\begin{tabular}{c c c}
%$\mathcal T_A$ & & $\mathcal T_B$ \\
%\hline
%%$\Tr(Q_i Q_j)$ & & $\Tr(q_i q_j)$ \\
%%$\Tr(B B)$ & & $\Tr(\phi_b)$ \\
%$\mathfrak M_A^{\cdot, 0}$ & & $\sigma_B$ \\
%$\mathfrak M_A^{0,\cdot}$ & & $\mathfrak M_B^{\cdot,\cdot}$ \\
%$\{{(\mathfrak {M}_A^{\cdot,\cdot}})_{Q_1^2}\}$ & & 
%$\{{(\mathfrak {M}_B^{0,\cdot}})_{Q_1^2}\}$ \\
%$\{{(\mathfrak {M}_A^{0,\cdot}})_{B^{2(N_2-1-k)}Q_1^{2k}  }\}$ & & 
%$\{{(\mathfrak {M}_B^{\cdot,\cdot}})_{A^{N_2-1-k}Q_1^{2k}}\}$
%%$\mathfrak M_A^{0, \cdot}$ & & $\mathfrak M_B^{\cdot, \cdot}$ \\
%\end{tabular}
%\ee

\subsubsection*{Second dual}
In order to dualize the $SU$ node we use the ARSW duality.

\paragraph*{ARSW duality $SU \leftrightarrow U$}
The duality that we need in order to dualize the $SU(2N_2)$ gauge node has been discussed in 
\cite{Aharony:2013dha} as the reduction of the original $SU(N)$ Seiberg duality \cite{Seiberg:1994pq}. Before going into the details of the second dual frame of theory $\mathcal T_A$ it is worth to review the original duality. 

The electric theory is an $SU(N)$ gauge theory with $F$ flavours $Q,\, \t Q$ and superpotential $\mathcal W_A=0$. The global symmetry is $SU(F) \times SU(F) \times U(1)_{A} \times U(1)_{B}$, where the baryonic symmetry $U(1)_B$ gives charges $+1$ to $Q$ and $-1$ to $\t Q$. The gauge invariant operators are the mesons $Q \t Q$, the baryons $Q^{N}$ and antibaryons $\t Q^N$. Moreover, we have the bare monopole operator $\mathfrak M_A$.

The magnetic theory is an $U(F-N)$ gauge theory with F flavours $q, \, \t q$, two fields $b, \, \t b$
charged only under the $U(1) \subset U(F-N)$ with charge $\pm (F-N)$, a matrix of $F \times F$ singlets $M$, and a singlet $Y$. The superpotential interactions are as follows 
\be
	\mathcal W_B= M q \t q + Y b \t b + \mathfrak M_B^{+}+ \mathfrak M_B^{-}.
\ee
Let us discuss the global symmetry of this theory. The non-abelian part is $SU(F) \times SU(F)$, while the abelian part in principle comprises the topological symmetry $U(1)_T$ apart from six independent $U(1)$'s acting on the chirals $M, \, q, \, \t q,\, Y, \, b, \, \t b$. Observe that being the gauge group $U$ and not $SU$ one combination of these $U(1)$'s is gauged; moreover the superpotential breaks four extra combinations, leaving only two abelian symmetries. Without going into the detail of the precise combination of the abelian symmetries, we just show the mapping of the gauge invariant operators that will be useful in the upcoming analysis\be
\begin{tabular}{c c c}
$\text{Electric}$ & & $\text{Magnetic}$ \\
\hline
$Q \t Q$ & & $M$ \\
$Q^N$ & & $q^{F-N} b$ \\
$\t Q^N$ & & $\t q^{F-N} \t b$ \\
$\mathfrak M_A$ & & $Y$ \\
\end{tabular}
\ee
where $Q^N=\epsilon_{i_1 \dots i_N} Q^{i_1}  \cdots  Q^{i_N}$, and similarly for $\t Q^N$; for simplicity we are suppressing flavour indices.
Observe that the linear monopole superpotential in the magnetic theory just removes from the chiral ring the monopoles for the $U(F-N)$ gauge theory.

Having discussed the duality involving an $SU(N)$ gauge theory we now come back to theory 
$\mathcal T_A$ and analyze the third dual frame: 
\be
\scalebox{1}{
\begin{tikzpicture}[baseline]
\tikzstyle{every node}=[font=\footnotesize]
%\node[draw, rectangle] (sqnodeL) at (-4,0) {$2$};
\node[draw=none] (node1) at (0,0) {$Sp(N_1)$};
\node[draw=none] (node2) at (3.5,0) {$U(2F+2N_1-2N_2)$};
\node[draw, rectangle] (sqnode1) at (7,0) {$2F$};
\node[draw, rectangle] (sqnode2) at (3.5,-2) {$2F+2N_1$};
\draw[black,solid,->] (node1) edge node[above]{$c$} (node2) ;
\draw[black,solid,<-] (node2) edge node[above]{$q_1$} (sqnode1) ;
\draw[black,solid,->] (node2) edge node[right]{$q_2$} (sqnode2) ;
\draw[black,solid,<-] (sqnode1) edge node[right]{$M$} (sqnode2) ;
\draw[black,solid,<-] (node1) edge node[left]{$N$} (sqnode2) ;

%\draw[black,solid] (node2) edge [out=45,in=135,loop,looseness=5]node[above]{$A$} (node2) ;
%\draw[black,solid,->] (node2) edge [out=25,in=155,loop,looseness=1]node[above]{$F$} (sqnodeR) ;
\node[draw=none] at (3.5,-3) {$\mathcal W=M q_1 q_2 + N c q_2+ Y b \t b+ \mathfrak M^{0, +} + \mathfrak M^{0, -} $};
\node[draw=none] at (-2.5,0) {$\mathcal T_{C}:$};
%node[above]{\blue $X$} node[below, yshift=0cm] {$\red F_X$} node {$ \red \times$}  (node);
\end{tikzpicture}}
\ee
Let us turn to the discussion on the chiral ring generators. We can start from the singlets flipping the mesons $q_1 q_2$ and $c q_2$, namely $M$ and $N$. However, $N$ is charged under the 
$Sp(N_1)$ gauge node, hence we need to take the gauge invariant combination $N^2$ where the indices of $Sp(N_1)$ are contracted using the symplectic tensor. Then, we have the singlet $Y$ that flips $b \t b$. The chiral ring also includes baryonic generators. The obvious one we can construct is using $q_2$ and $b$: $q_2^{2F+2N_1-2N_2} \tilde b$, where as explained in the previous section the presence of $\tilde b$ is necessary to balance the $U(1)$ gauge charge. This operator transforms in the rank-$(2N_2)$ antisymmetric representation of $SU(2F+2N_1)$ global symmetry.
There is another baryon we may construct, this time using three chirals: $q_1, \, c, \, b$ contracted as $q_1^{2F-2N_2+2k} \, c^{2N_1-2k} \, b$ and transforms in the rank-$(2N_2-2k)$ antisymmetric of $SU(2F)$.
The monopoles with flux only on the $U(2F+2N_1-2N_2)$ enters linearly in the superpotential and are removed from the chiral ring. 
The monopole operators of the form $\mathfrak M_C^{\cdot, 0}$ needs a detailed discussion. Usually, if we take a $Sp(N)$ gauge theory with $2F$ fundamental chirals, the basic monopole operator is charged under the axial symmetry (in particular it has charge $-2F$.) In theory $\mathcal T_C$ this axial symmetry is gauged because of the $U(2F+2N_1-2N_2)$ gauge node, hence the basic $Sp(N_1)$ monopole operator is not gauge invariant. To construct a gauge invariant operator out of the $Sp(N_1)$ monopole it is possible to dress it using the fields with the fields $q_1$ and $c$, which have the correct $U(1)$ gauge charge to cancel the one of the monopole. The reason is the following: the axial charge of $\mathfrak M_C^{\cdot, 0}$, which is $-(2F+2N_1-2N_2)$, has opposite sign with respect to the $U(1)$ charge of $c$, that we take to be $+1$. Moreover, we see from the quiver $\mathcal T_C$ that $q_1$ has the same $U(1)$ charge as $c$. Thus, the $U(1)$ gauge charge of the bare $\mathfrak M_C^{\cdot, 0}$ is cancelled if we dress with a total of $2F+2N_1-2N_2$ fields, being either $c$ or $q_1$. Recall that we need to make the dressed monopole gauge invariant also under $Sp(N_1)$ and $SU(2F+2N_1-2N_2) \subset U(2F+2N_1-2N_2)$: this is achieved by taking an even number of $c$ fields and contracting everything with the epsilon tensor of $U(2F+2N_1-2N_2)$. The tower of dressed monopoles constructed in this way is as follows
\be
	\{(\mathfrak M_C^{\cdot, 0})_{q_1^{2F+2N_1-2N_2-2k} c^{2k}} \},
\ee
where $q_1^{2F+2N_1-2N_2-2k} c^{2k}$ is a schematic expression denoting the baryonic operator
\be
	(q_1)_{i_1} \cdots (q_1)_{i_{2F+2N_1-2N_2-2k}} c_{i_{2F+2N_1-2N_2-2k+1}} \cdots 
	c_{i_{2F+2N_1-2N_2}}
\ee

%\be
%\begin{tabular}{c c c}
%$\mathcal T_A$ & & $\mathcal T_C$ \\
%\hline
%$B BQ_2 Q_2$ & & $NN$ \\
%$Q_2 Q_1$ & & $M$ \\
%$Q_2^{2N_2}$ & & $q_2^{2F+2N_2-2N_1} \t b$ \\
%%$\Tr(B B)$ & & $\Tr(\phi_b)$ \\
%%$\mathfrak M_A^{\cdot, 0}$ & & $\sigma_B$ \\
%$\mathfrak M_A^{0,\cdot}$ & & $Y$ \\
%$\mathfrak M_A^{\cdot,0}$ & & $\{\mathfrak M_C^{\cdot,0} \,b\}$ \\
%$\{{(\mathfrak {M}_A^{\cdot,\cdot}})_{Q_1^2}\}$ & & 
%$\{{(\mathfrak {M}_C^{\cdot,0}})_{q_1^2}\}$ \\
%\end{tabular}
%\ee

\subsubsection*{Operator map}
Finally we are ready to discuss the map of chiral ring generators across the three dual frames of the $SU-Sp$ gauge theory we studied. 
As we have done in the other sections, it is useful to have at hand the $R$-charge map that allows to go from $\mathcal T_A$ to $\mathcal T_B$ and $\mathcal T_C$:

\begin{itemize}
\item{$\mathcal T_B \to \mathcal T_A$:
\be	
	R_b=1-R_B, \qquad R_{A}=2R_B.
\ee
}
\item{$\mathcal T_C \to \mathcal T_A$:
\begin{align}	
	R_{q_1}&=\frac{2N_2(R_{Q_1}-1)+N_1(2+R_B-2R_{Q_1}-R_{Q_2})-F(R_{Q_1}+R_{Q_2}-2)}{2F+2N_1-2N_2} \\
	R_{q_2}&=\frac{2N_2(R_{Q_2}-1)-N_1(-2+R_B+R_{Q_2})-F(R_{Q_1}+R_{Q_2}-2)}{2F+2N_1-2N_2} \\
	R_{b}&=R_{\tilde b}=2N_2+N_1(R_B+R_{Q_2}-2)+F(R_{Q_1}+R_{Q_2}-2) \\
	R_{c}&=\frac{2N_2(R_{B}-1)+F(2-2R_B+R_{Q_1}-R_{Q_2})-N_1(R_{Q_2}+R_{B}-2)}{2F+2N_1-2N_2}
\end{align}
}
\end{itemize}
The operator map is as follows
\be
\begin{tabular}{c c c c c}
$\mathcal T_A$ & & $\mathcal T_B$ & & $\mathcal T_C$ \\
\hline
$B^2 Q_2^2$ & & $A Q_2 Q_2$ & & $N^2$ \\
$Q_2 Q_1$ & & $Q_2 Q_1$ & & $M$ \\
$B^{2k}Q_1^{2N_2-2k}$ & & $A^k Q_1^{2N_2-2k}$ & & $q_1^{2F-2N_2+2k}\,c^{2N_1-2k}\,b$ \\
$Q_2^{2N_2}$ & & $Q_2^{2N_2}$ & & $q_2^{2F+2N_1-2N_2} \t b$ \\
$\mathfrak M_A^{0,\cdot}$ & & $\mathfrak M_B^{\cdot,\cdot}$ & & $Y$ \\
$\mathfrak M_A^{\cdot,0}$ & & $\sigma_B$ & & $\{\mathfrak M_C^{\cdot,0} \,b\}$ \\
$\{{(\mathfrak {M}_A^{\cdot,\cdot}})_{Q_1^{2k}B^{2(N_2-1-k)}}\}$ & & 
$\{{(\mathfrak {M}_B^{0,\cdot}})_{Q_1^{2k} A^{N_2-1-k}}\}$ & & 
$\{{(\mathfrak {M}_C^{\cdot,0}})_{q_1^{2(F-k)} c^{2(k+N_1-N_2)}}\}$ \\
\end{tabular} 
\ee
Observe the last two lines of the table which involves dressed monopoles. The $Sp(N_1)$ monopole $\mathfrak M_A^{\cdot, 0}$, mapping to the usual Aharony singlet $\sigma_B$ in $\mathcal T_B$, maps to a dressed monopole in theory $\mathcal T_C$: 
$\{ \mathfrak M_C^{\cdot, 0} \, b \}$. The last line of the table is similar to the mapping of the monopole with flux on two nodes in the case of $Sp-Sp$ gauge theory, the difference is that now we have non trivial dressing making the monopoles transforming in the rank-$2k$ antisymmetric representation of the $SU(2F)$ global symmetry group. 

\subsubsection*{Supersymmetric index}
We compute the supersymmetric index for the triality at hand in the case of $N_1=1, \, N_2=2, \, F=2$ with the choice of $R$-charges given by $R_B=1/5$, $R_{Q_1}=3/8$, $R_{Q_2}=4/7$. 
For theory $\mathcal T_B$ we did not manage to perform the index computation due to machine limitations, so we limited ourselves to the explicit computation for $\mathcal T_A$ and $\mathcal T_C$ theories. 

\be
\begin{split}
\mathcal I&=1+{\color{blue}{24 x^{45/56} y_{Q_1} y_{Q_2}}}+{\color{blue}{6 x^{23/20} y_B^2 y_{Q_1}^2}}+{\color{blue}{\frac{x^{6/5}}{y_B^4}}}+{\color{blue}{15 x^{44/35} y_B^2 y_{Q_2}^2}}+\\
&+{\color{blue}{x^{3/2} y_{Q_1}^4}}+{\color{blue}{\frac{x^{107/70}}{y_B^2 y_{Q_1}^4 y_{Q_2}^6}}}+300 x^{45/28} y_{Q_1}^2 y_{Q_2}^2+{\color{blue}{15 x^{12/7} y_{Q_2}^4}}+
{\color{blue}{\frac{6 x^{263/140}}{y_B^4 y_{Q_1}^2 y_{Q_2}^6}}}+\\
&+144 x^{547/280} y_B^2 y_{Q_1}^3 y_{Q_2}-53 x^2+\dots
\end{split}
\ee
Finally we can identify the various generators in the index 
\begin{align}
	\Tr(B^2 {Q_2}^2) \;&\leftrightarrow\; {\color{blue}{15 x^{44/35} y_B^2 y_{Q_2}^2}}\\
	\Tr(Q_2 Q_1) \;&\leftrightarrow\; {\color{blue}{24 x^{45/56} y_{Q_1} y_{Q_2}}}\\
	\Tr({Q_1}^4) \;&\leftrightarrow\; {\color{blue}{x^{3/2} y_{Q_1}^4}}\\
	\Tr(B^2 {Q_1}^2) \;&\leftrightarrow\; {\color{blue}{6 x^{23/20} y_{B}^2 y_{Q_1}^2}}\\
	\Tr({Q_2}^4) \;&\leftrightarrow\; {\color{blue}{15 x^{12/7} y_{Q_2}^4}}\\
	\mathfrak M_A^{0, \cdot}\;&\leftrightarrow\; {\color{blue}{x^{107/70} 
	\frac{1}{y_{B}^2 y_{Q_1}^4 y_{Q_2}^6}}} \\
	\mathfrak M_A^{\cdot, 0}\;&\leftrightarrow\; {\color{blue}{x^{6/5} 
	\frac{1}{y_{B}^4}}} \\
	\{(\mathfrak M_A^{\cdot, \cdot})_{Q_1^2} \}\;&\leftrightarrow\; {\color{blue}{x^{263/140} 
	\frac{6}{y_{B}^4 y_{Q_1}^2y_{Q_2}^6}}} \\
\end{align}

\newpage
\section{Conclusion and open questions}
In this note we discussed Seiberg-like dualities for two-node quiver theories with various gauge groups, paying particular attention to the mapping of monopole operators across the duality. We may identify two different class of theories: one class without baryonic operators, such as quivers involving two unitary or symplectic groups, and one class with baryons and baryon-monopoles, such as various combination of special orthogonal, special unitary and symplectic groups. Even though we explicitly studied only two-node quivers, for the first class we find a simple rule to map all chiral ring generators (mesons and monopoles) in a quiver of arbitrary length. We were not able to find such a general rule for the second class of theories. In this case there are subtle issues, especially for orthogonal groups, that do not appear for unitary or symplectic quivers that makes finding such a general map a hard task. Nonetheless, the chiral-ring in the presence of baryons and baryon-monopoles is much richer and involves interesting operators that are not present when discussing single node quivers. 

A very useful tool in our analysis is the superconformal index, whose explicit expression (as a function of the fugacities for the abelian global symmetries, turning on fugacities for non-abelian symmetries is computationally hard and not necessary for our purposes), allows us to check that our proposed list of chiral ring generators is complete.

The analysis performed here has its own interest as a step towards the understanding of the complete operator map across duality in general $3d$ $\mathcal N=2$ quivers. Nonetheless, our first motivation was to use the techniques developed here to study the ``deconfinement" of rank-2 matter, for instance in the case of an adjoint $U(N)$ chiral or for the rank-2 anti-symmetric representation for $Sp(N)$. Aspects of the latter theory have been discussed in \cite{Benvenuti:2018bav, Amariti:2018wht}. These results will be presented in \cite{BGLMseqdec:2020,BGLMseqdec:2020b}.

Finally, let us mention some open questions.

 There are some two-node quivers for which we have not been able to understand the complete chiral-ring map, such as for $Sp(N_1) \times U(N_2)$ and $SU(N_1) \times U(N_2)$, where the matter fields are such that the full theory is non-chiral. It would be interesting to fully understand these cases. 
The last comment immediately raises the question of analizing theories with chiral matter content. The difficulty in such cases comes already from the intricate structure of the dualities needed for chiral theories. Some results on dualities for chiral-matter are found in \cite{Aharony:2014uya, Nii:2018bgf, Nii:2019qdx, Nii:2020ikd, Amariti:2020xqm}.
Furthermore, a more systematic analysis of alternating orto-symplectic quivers is desirable, in the spirit of theories with eight supercharges. 
All the developments we mentioned up now involve theories with four supercharges. However, another interesting line of research is related to theories with minimal amount of supersymmetry, $\mathcal N=1$ in $2\!+\!1$ dimensions. Some previous works on the dynamics of such theories can be found in \cite{Bashmakov:2018wts, Benini:2018umh, Eckhard:2018raj, Gaiotto:2018yjh, Benini:2018bhk, Choi:2018ohn, Fazzi:2018rkr, Rocek:2019eve, Aharony:2019mbc, Bashmakov:2018ghn, Bashmakov:2019myq,   Sharon:2020xod}. To the best of our knowledge there is no analysis for the IR dynamics of $\mathcal N=1$ quivers and it would be worth studying it.

\acknowledgments
We are grateful to Antonio Amariti, Noppadol Mekareeya, Sara Pasquetti and Matteo Sacchi for valuable discussions. GLM is supported by the Swedish Research Council grant number 2015-05333 and partially supported by ERC Consolidator Grant number 772408 ``String landscape''.

%%%%%%%%%%%%%%%%%%%%%%%%%%%%%%%%%%%%%%%%%%%%%%%%%%%%%%%%%%%%%%%
%%%%%%%%%%%%%%%%%%%%%%%%%%%%%%%%%%%%%%%%%%%%%%%%%%%%%%%%%%%%%%%
%%%%%%%%%%%%%%%%%%%%%%%%%%%%%%%%%%%%%%%%%%%%%%%%%%%%%%%%%%%%%%%

\appendix

\section{$3d$ Supersymmetric index} \label{appA}
The $3d$ supersymmetric index is a powerful tool allowing to analize various properties and dualities for theories with at least four supercharges. Its power lies on being an RG-invariant quantity, thus one can study the properties of a strongly coupled fixed point via the weak coupling description of the theory under study.
As usual for the Witten index, the definition involves a trace over the Hilbert space of the theory on 
$S^2 \times \mathbb R$ \cite{Bhattacharya:2008zy, Bhattacharya:2008bja, Kim:2009wb, Imamura:2011su, Kapustin:2011jm, Dimofte:2011py}, (we use the definitions of \cite{Aharony:2013dha, Aharony:2013kma}):
\be\label{TrInd}
	\mathcal I(x, \vec\mu)=\Tr \left[ (-1)^{J_3} x^{\Delta + J_3} \prod_{i} \mu_i^{q_i} \right],
\ee
where the various quantities in the formula represents
\begin{itemize}
	\item{$\Delta$: is the energy whose scale is set by the radius of $S^2$,}
	\item{$J_3$: is the Cartan generator for the $SO(3)$ isometry of the $S^2$,}
	\item{$\mu_i, \, q_i$: respectively the fugacities and charges of the global non-$R$ 			symmetries.}
\end{itemize}
The only non trivial contributions that enter the index comes from states annihilated by two supercharges and satisfy the following condition
\be
	\Delta=R+J_3,
\ee
$R$ being the $R$-charge.

It is not so easy to employ the definition \eqref{TrInd} to perform an explicit computation of the index; here the localization techniques come at rescue. Indeed, the index can be computed as the partition function on $S^2 \times S^1$ given by the following expression
\be\label{LocInd}
	\mathcal I(x)=\sum_{\bold m} \frac{1}{|\mathcal W_{\bold m}|} \int
	\frac{d \bold z}{2 \pi i \bold z} Z_{\text{cl}} \, Z_{\text{vec}} \, Z_{\text{mat}},
\ee
where the integral is taken over the Cartan torus of the gauge group whose fugacities are $\bold z$, $|\mathcal W_{\bold m}|$ is the dimension of the Weyl group that is left unbroken by the monopole background specified by the GNO magnetic fluxes $\bold m$. Localization implies that the only non trivial contribution to \eqref{LocInd} from non-exact term in the classical action and  
from $1$-loop terms. The various terms $Z_{\text{cl}}, \, Z_{\text{vec}}, \, Z_{\text{mat}}$ have the following expressions
\begin{itemize}
	\item{$Z_{\text{cl}}$: The classical terms includes only CS couplings and, more generally, BF 	terms. Take a gauge group whose rank is $\text{rk}G$. Denoting the fugacity for the 		topological with $\omega$ and the associated flux as $\bold n$, and given a level $k$ CS term we have
	\be
		Z_\text{cl}=\prod_{i=1}^{\text{rk}G} \omega^{m_i} z_i^{k\, m_i+\bold n}
	\ee}
	\item{$Z_{\text{vec}}$: The contribution for an $\mathcal N=2$ vector multiplet reads
	\be
		Z_{\text{vec}}=\prod_{\alpha \in \mathfrak g} x^{-\frac{|\alpha(m)|}{2}} (1-(-1)^{\alpha(m)}
		\bold z^\alpha x^{|\alpha(m)|})
	\ee}
	\item{$Z_{\text{mat}}$: The contribution of an $\mathcal N=2$ chiral multiplet with $R$-charge $r$ transforming in the representations $\mathcal R$ and $\mathcal R_F$ under the gauge and flavour group, whose weights we denote as $\rho, \rho_F$, is
	\begin{align}
		Z_{\text{chi}}=\prod_{\rho \in \mathcal R} \prod_{\rho_F \in \mathcal R_F} 
		&(z^\rho \mu^{\rho_F} x^{r-1})^{-\frac{|\rho(m)+\rho_F(n)|}{2}} \times \\
		&\times \frac{((-1)^{\rho(m)+\rho_F(n)} z^{-\rho} \mu^{-\rho_F} x^{2-r+|\rho(m)+\rho_F(n)|; x^2})_\infty}{((-1)^{\rho(m)+\rho_F(n)} z^{-\rho} \mu^{-\rho_F} x^{r+|\rho(m)+\rho_F(n)|; x^2})_\infty}.
	\end{align}}
\end{itemize}

\section{Monopoles and dualities for orthogonal gauge groups}
\label{sec:reviewSO}
In this section we will review the current knowledge about the monopole operators in theories involving orthogonal gauge groups and the related Seiberg-like dualities proposed in \cite{Aharony:2011ci,Aharony:2013kma}. 

Let us start considering an $SO(N)$ theory with $F$ flavors $Q$. The global symmetry group of this model is:
\be
\label{eq:SOglobalsym}
G_{N,F}\,=\,(U(F)\times \mathbb{Z}^\cC_2\times \mathbb{Z}_2^\cM)/\mathbb{Z}_2\,,
\ee
where the Abelian factor in $U(F)$ is the $U(1)_Q$ axial symmetry acting on the chiral fields; the discrete $\mathbb{Z}_2^\cC$ factor is the charge conjugation symmetry, whose non-trivial element consists of the orthogonal transformation (in $O(N)$) with determinant equal to $-1$, {\it i.e.} a reflection; the magnetic discrete symmetry $\mathbb{Z}_2^\cM$, instead, acts on the Coulomb branch coordinates charging $-1$ the fundamental monopole operators.\footnote{The two discrete factors $\mathbb{Z}_2^{\cC,\cM}$ and the element $e^{i\pi}$ of $U(1)_Q$ are not really independent but they actually satisfy the relation $e^{i\pi}_Q\cdot \cC^N\cdot \cM^F=1$ \cite{Aharony:2013kma}; this is the reason why a common $\mathbb{Z}_2$ factor is mod out in \eqref{eq:SOglobalsym}.} As usual, on the Coulomb branch the gauge group is broken to the Cartan $U(1)^{r_N}$ with $r_N=\lfloor N/2\rfloor$. Semi-classically, the basic monopole operators can be written as:
\be
\cV_{\pm}\,\approx\,e^{\pm\left(\tfrac{\alpha_1}{g^2}+i\phi_1\right)}\,,
\ee
where we denoted with $\alpha_i$ and $\phi_i$ the dual photon and adjoint scalar respectively for the $i^{th}$ Abelian vector multiplet in $U(1)^{r_N}$. Charge conjugation acts non-trivially on the two monopoles $\cV_{\pm}$ swapping them, so that it is useful to define the even and odd $\mathbb{Z}_2^\cC$ combinations
\be
\mathfrak M^{\pm}\,=\,\cV_{+}\,\pm\, \cV_-\,.
\ee
Observe that both the monopoles breaks the gauge group down to $S\left(O(N-2)\times O(2)\right)$, including the transformation with $-1$ determinant in both the $O(N-2)$ and the $O(2)$ factors. In particular, in order for the monopole to be gauge invariant, it must be invariant under charge conjugation in the $O(2)$ factor; following the previous discussion, only $\mathfrak{M}^+$ has this property, while $\mathfrak M^-$ is not gauge invariant on its own. 

However, we can still build a gauge invariant object dressing the monopole with an operator that is odd with respect to the charge conjugation in $SO(N-2)$:
\be
(\mathfrak M^-)_{Q^{N-2}}\,\approx\, \mathfrak{M}^- \cdot \varepsilon_{i_i\dots i_{N-2}}Q^{i_1}\cdots Q^{i_{N-2}}\,
\ee
where the chiral fields are contracted using the Levi-Civita symbol of the residual $SO(N-2)$ factor, $\varepsilon$. This monopole operator is usually called \emph{baryon monopole}: let us observe that it only exists for $F\geq N-2$, it has non-trivial charge under both $\mathbb{Z}_2^\cC$ and $\mathbb{Z}_2^\cM$ and transforms in the rank-$(N-2)$ antisymmetric representation of $SU(F)$.

Another type of operator is relevant for us, having non-trivial magnetic fluxes with respect to two different Abelian factors in $U(1)^r$. Semi-classically, it can be written as:
\be
\mathfrak M^{\div}\,\approx\, \text{exp}\left(\frac{\alpha_1-\alpha_2}{g^2}+i(\phi_1-\phi_2)\right)\,.
\ee
where the two lined up bullets denote the fact that two different fluxes are turned on. Such monopole breaks the $SO(N)$ gauge group down to $S(O(N-4)\times O(4))$,\footnote{Actually, the $SO(4)$ factors is further broken to $U(2)$.} and it is not gauge invariant unless dressed with a conjugation-odd operator in $SO(N-4)$. This can be done defining the gauge invariant operator $\varepsilon\mathfrak M^\div_{Q^{N-4}}$,\footnote{The $N-4$ chiral fields dressing the monopoles are contracted with the Levi-Civita symbol of the $SO(N-4)$ residual group.} existing only for $F\geq N-4$. In theories with only one gauge group factor, $\mathfrak M^{\div}$ is not really a chiral operator; however, it plays a crucial role in dualities between orthogonal quiver theories discussed in this paper. 

In the theory under consideration, the last operator that deserve to be mentioned is the usual baryon:
\be
\label{eq:BaryonDef}
\cB\,=\, \varepsilon_{i_1\dots i_N}\,Q^{i_1}\cdots Q^{i_N}\,.
\ee
In the main text, different baryon-like operators can appear; in that case, we will denote them by an $\varepsilon$ followed by the fields contracted with the Levi-Civita symbol: for instance, the baryon in \eqref{eq:BaryonDef} could be also denoted by $\varepsilon Q^N$.

Once we have understood which kind of operators can be part of the chiral ring in 3d SQCD with orthogonal groups, we can easily discuss the Seiberg-like duality proposed by Aharony, Razamat, Seiberg and Willett (ARSW) in \cite{Aharony:2013kma}. The theory dual of $\CT_A$, $SO(N)$ SQCD with $F$ flavors $Q$, $\CW_{\CT_A}=0$, is $\CT_B$,  $SO(F-N+2)$ gauge theory with $F$ flavors $q$, $F(F+1)/2$ singlets $M_{ij}$ transforming in the symmetric representation of $SU(F)$ and superpotential:
\be
\cW_{\CT_B}\,=\, \sigma \mathfrak{M}^+\,+\, M_{ij}{\Tr} (q^iq^j)\,.
\ee
The map of the chiral ring generators is the following:
\be
\begin{tabular}{c c c }
$\mathcal T_A$ & & $\mathcal T_B$ \\
\hline
$\Tr(Q_i Q_j)$ & & $M_{ij}$  \\
$\mathfrak M^+$ & & $\sigma$  \\
$(\mathfrak M^-)_{Q^{N-2}}$ & & $\varepsilon q^{N-F+2}$  \\
$\varepsilon Q^{N}$ & & $(\mathfrak M^-)_{q^{N-F}}$  \\
\end{tabular}
\ee
Observe that baryons and baryon-monopoles are mapped to each other.

\subsection{$O(N)_\pm\,,\,\mathrm{Pin}(N)\text{ and }\mathrm{Spin}(N)$}
Different gaugings of charge conjugation and the magnetic $\mathbb{Z}_2^\cM$ discrete symmetry leads to different gauge groups, enjoying the same algebra as $SO(N)$ but differing in their global properties; in particular, the spectrum of chiral operators will be different.
\begin{itemize}
\item The gauge group $O(N)_+$ is obtained gauging $\mathbb{Z}_2^\cC$, {\it i.e.} the orthogonal reflection. Such $O(N)$ group is the most common in literature: the gauging of charge conjugation makes the baryon and the baryon-monopole not gauge invariant and they are not part of the chiral ring anymore.

\item If we gauge the diagonal combination $(\mathbb{Z}_2^\cC\times \mathbb{Z}_2^\cM)/\mathbb{Z}_2$, the less common $O(N)_-$ group is obtained; in this theory, only operators which are even (odd) under both charge conjugation and $\mathbb{Z}^\cM_2$ symmetry are gauge invariant: for this reason the monopole $\mathfrak{M}^+$ and the baryon $\cB$ are both projected out, while the baryon-monopole $(\mathfrak M^-)_{ Q^{N-2}}$ survives. However, the monopole usually denoted as $\mathfrak{M}^+_{\mathrm{Spin}}$, having twice the minimal flux, survives. 

\item $\mathrm{Spin}(N)$ theories are built gauging $\mathbb{Z}_2^\cM$. The (baryon-)monopole is projected out but the monopoles with double fluxes, $\mathfrak M^+_{\mathrm{Spin}}$ and $(\mathfrak M^-_{\mathrm{Spin}})_{Q^{N-2}}$, are still chiral operators on the Coulomb branch.

\item Finally, $\text{Pin}(N)$\footnote{To be precise, there exists two versions of $\text{Pin}(N)$: 
$\text{Pin}^\pm(N)$, as discussed in \cite{Cordova:2017vab}.} theories are obtained gauging both the discrete global symmetries; the Coulomb branch is parametrized by $\mathfrak{M}_{\mathrm{Spin}}^+$ while all baryonic-like operators (including monopoles) are projected out.
\end{itemize}

All such theories enjoy Seiberg-Like duality similar to the ARSW duality \cite{Aharony:2011ci,Aharony:2013kma}. $O(N)_+$ SQCD with $F$ flavors is dual to $O(F-N+2)_+$ SQCD with $F$ flavors, $N(N+1)/2$ $M_{ij}$ singlets duals of the meson ${\Tr} Q_i Q_j$, the singlet $\sigma$ dual of $\mathfrak M^+$ and the usual superpotential $\cW=\sigma \mathfrak M^++{\Tr}(q^i M_{ij}q^j)$; an analogous duality holds for $\text{Pin}(N)$ SQCD. Finally, $O(N)_-$ SQCD is dual to $\text{Spin}(F-N+2)$ SQCD (with singlets and appropriate superpotential): further details about the chiral ring mapping can be found in \cite{Aharony:2013kma}.
\bibliographystyle{ytphys}
\bibliography{ref}
\end{document}